\documentclass[preprint2]{aastex}

\slugcomment{DRAFT: \today} 

\newcommand{\Teff}{\hbox{$T_{\rm eff}$}} \def\logg{$\log g$\thinspace} \def\etal{{et\thinspace al.}\ }  \def\md{log($\dot{M}$/M$_{\odot} {\rm yr}^{-1}$)\,}
\newcommand{\lppr}{\stackrel{<}{\scriptstyle \sim}}
\newcommand{\lappr}{\raisebox{-0.4ex}{$\lppr $}}
\newcommand{\gppr}{\stackrel{>}{\scriptstyle \sim}}
\newcommand{\gappr}{\raisebox{-0.4ex}{$\gppr $}}
\newcommand{\msun}{\ensuremath{\, {\rm M}_\odot}}
\newcommand{\teff}{\ensuremath{T_{\rm eff}}}  \newcommand{\lsun}{\ensuremath{\,{\rm L}_\odot}} \newcommand{\abb}[1]{Fig.\,\ref{#1}}
\newcommand{\kap}[1]{Sect.\,\ref{#1}}
\newcommand{\kaps}[1]{Sections~\ref{#1}}
\newcommand{\czw}{\ensuremath{^{12}\mem{C}}}
\newcommand{\nezw}{\ensuremath{^{22}\mem{Ne}}}
\newcommand{\n}{\ensuremath{\mem{n}}}  \newcommand{\p}{\ensuremath{\mem{p}}}
\newcommand{\mgfu}{\ensuremath{^{25}\mem{Mg}}}
\newcommand{\lisi}{\ensuremath{^{7}\mem{Li}}}
\newcommand{\besi}{\ensuremath{^{7}\mem{Be}}}
\newcommand{\fese}{\ensuremath{^{56}\mem{Fe}}}
\newcommand{\hedr}{\ensuremath{^{3}\mem{He}}}
\newcommand{\emi}{\ensuremath{\mem{e}^\mem{-}}}
\newcommand{\jahre}{\ensuremath{\, \mathrm{yr}}}
\newcommand{\mem}[1]{\ensuremath{\mathrm{ #1}}}
\newcommand{\ose}{\ensuremath{^{16}\mem{O}}}
\newcommand{\oac}{\ensuremath{^{18}\mem{O}}}
\newcommand{\ndr}{\ensuremath{^{13}\mem{N}}}
\newcommand{\cdr}{\ensuremath{^{13}\mem{C}}}
\newcommand{\cvi}{\ensuremath{^{14}\mem{C}}}
\newcommand{\siac}{\ensuremath{^{28}\mem{Si}}}
\newcommand{\sine}{\ensuremath{^{29}\mem{Si}}}
\newcommand{\sidr}{\ensuremath{^{30}\mem{Si}}}
\newcommand{\spr}{\mbox{$s$-process}}  

\newcommand{\hevi}{\ensuremath{^{4}\mem{He}}}
\newcommand{\nvi}{\ensuremath{^{14}\mem{N}}}
\newcommand{\nfu}{\ensuremath{^{15}\mem{N}}}

\newcommand{\fne}{\ensuremath{^{19}\mem{F}}}
\newcommand{\kelv}{\ensuremath{\,\mathrm K}}

\shorttitle{Bare PN Central Stars} \shortauthors{Werner \& Herwig}

\begin{document}

\title{The Element Abundances in Bare Planetary Nebula Central Stars and the Shell Burning in AGB
Stars}  \author{Klaus Werner} \affil{Institut f\"ur Astronomie und Astrophysik,
Universit\"at T\"ubingen, Sand~1, D-72076 T\"ubingen, Germany}
\email{werner@astro.uni-tuebingen.de} \and \author{Falk Herwig} \affil{Los
Alamos National Laboratory, Theoretical Astrophysics Group T-6, MS B227, Los
Alamos, NM 87545, U.S.A.} \email{fherwig@lanl.gov}

\begin{abstract}
We review the observed properties of extremely hot hydrogen-deficient post-AGB
stars of spectral type [WC] and PG1159. Their H-deficiency is probably caused by
a (very) late helium-shell flash or a AGB final thermal pulse, laying bare interior stellar regions which are usually
kept hidden below the hydrogen envelope. Thus, the photospheric element
abundances of these stars allow to draw conclusions about details of nuclear
burning and mixing processes in the precursor AGB stars. We summarize the
state-of-the-art of stellar evolution models which simulate AGB evolution and
the occurrence of a late He-shell flash. We compare predicted element abundances
to those determined by quantitative spectral analyses performed with advanced non-LTE
model atmospheres. A good qualitative and quantitative agreement is found. 
Future work can contribute to an even more complete picture of the
nuclear processes in AGB stars.
\end{abstract}


\section{Introduction}

\begin{figure*}
\begin{center}
\includegraphics[scale=.60]{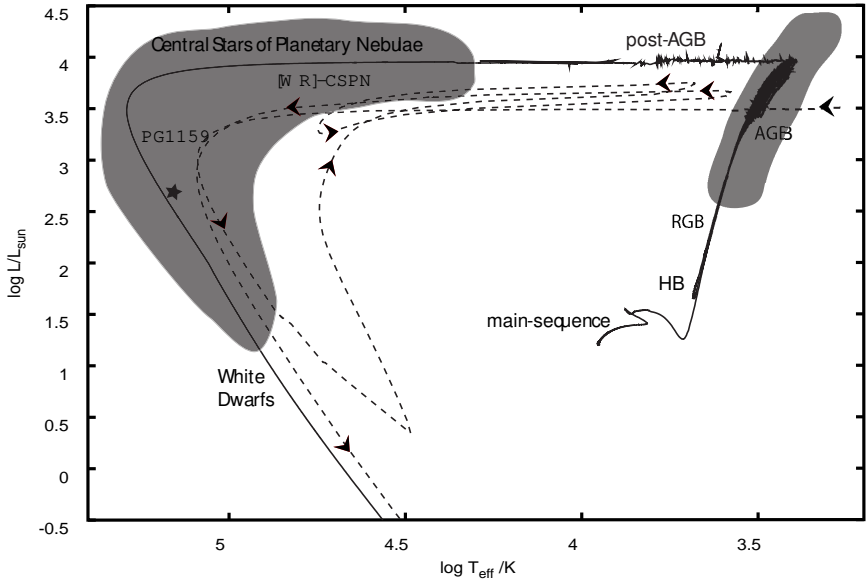}
\end{center}
\figcaption{\label{fig:hrd} Complete stellar evolution track with an initial mass
of $2\msun$ from the main sequence through the Red Giant Branch phase, the
Horizontal Branch phase to the Asymptotic Giant Branch phase, and finally
through the post-AGB phase that includes the central stars of planetary nebulae
to the final white dwarf stage. The solid line represents the evolution of a
H-normal post-AGB star. The dashed line shows a born-again evolution of the same
mass, triggered
by a very late thermal pulse, however, shifted by
approximately $\Delta \log \teff = - 0.2$ and $\Delta \log L/\lsun = - 0.5$ for
clarity. The star shows the position of PG1159-035. }
\end{figure*}                                                                                        

\begin{figure*}
\begin{center}
\includegraphics[scale=.60]{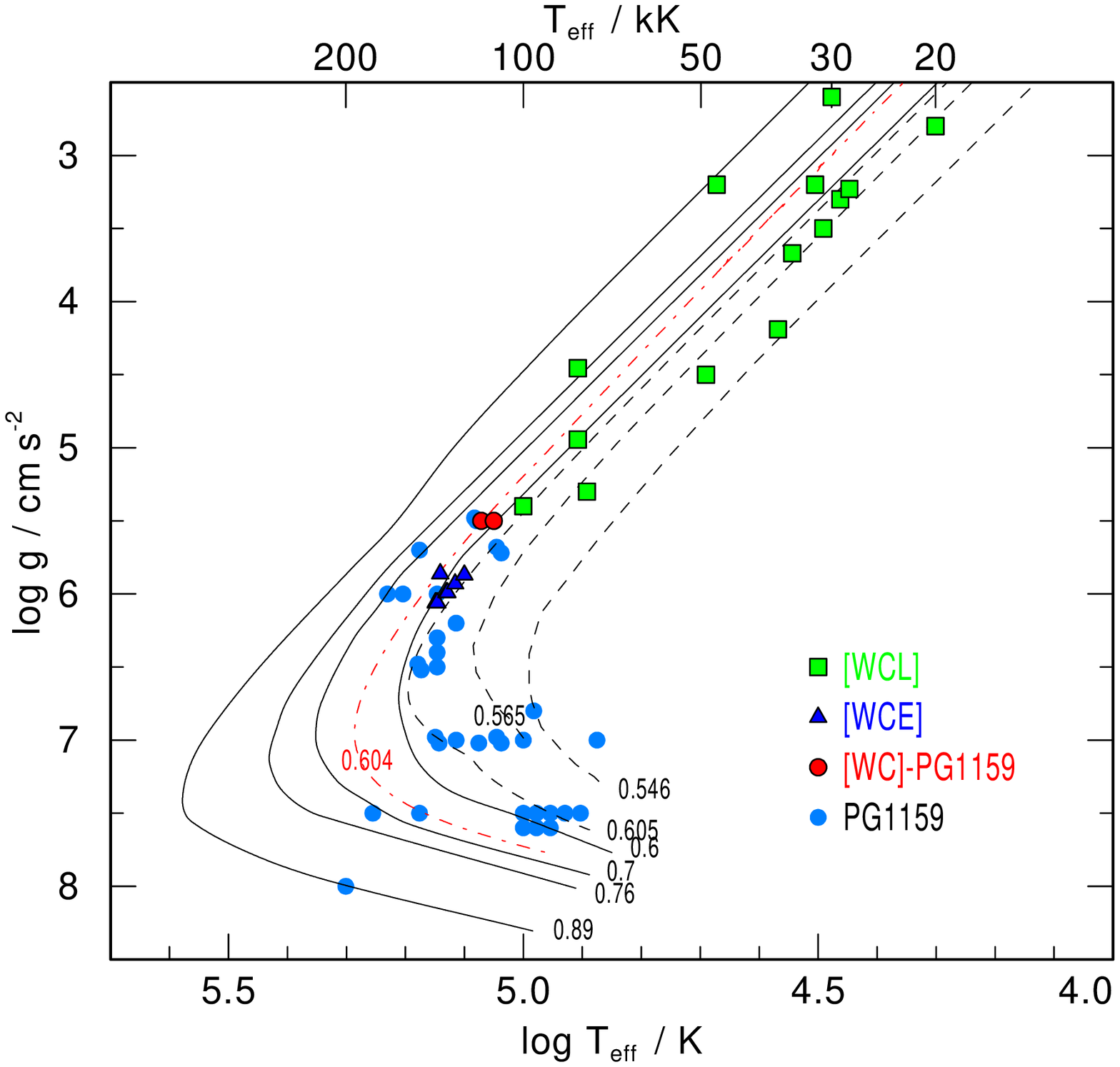}
\end{center}
\caption{Hot hydrogen-deficient post-AGB stars in the
$g$--\Teff--plane. We identify Wolf-Rayet central stars of early and
late type \citep[{[WCE], [WCL],} from][]{hamann:97}, PG1159 stars (from
Table~\ref{tabpg1159}) as well as two [WC]--PG1159 transition objects
(Abell~30 and 78). \Teff\ for the [WC] stars is related to the
stellar radius at $\tau_{\rm Ross}$=20. Evolutionary tracks are from
\citet{schoenberner:83} and \citet{bloecker:95b} (dashed lines),
\citet{wood:86} and \citet{2003IAUS..209..111H} (dot-dashed line)
(labels: mass in M$_\odot$).  The latter $0.604\msun$ track is the
final CSPN track following a VLTP evolution and therefore has a
H-deficient composition. However, the difference between the tracks is
mainly due to the different AGB progenitor evolution.}\label{fighrd}
\end{figure*}

Post-AGB stars represent a relatively short-lived transition phase between
Asymptotic Giant Branch (AGB) stars and white dwarf (WD) stars. All stars with
initial masses between 1 and about 8~M$_\odot$ become H- and He-shell burning
AGB stars and end their lives as WDs with a carbon-oxygen core. The model
evolution is shown in \abb{fig:hrd} for initially $2\msun$.  More massive
Super-AGB stars up to $12\msun$ may end as ONeMg WDs.  During the hottest
phases, post-AGB stars are surrounded by a planetary nebula (PN), ionized material lost by the precursor
AGB star. Canonical stellar evolution predicts that throughout all evolutionary
phases these stars retain hydrogen-rich envelopes which, however, can become
contaminated by processed material from the interior by dredge-up events
occurring in the Red Giant Branch (RGB) and AGB stages. Such an evolution is described
by the solid line in \abb{fig:hrd}.

Already decades ago, the observation of central stars of planetary nebulae
(CSPN) exhibiting emission-line spectra which are very similar to those of
massive Wolf-Rayet stars with strong helium and carbon emission lines
(i.e. spectral type WC) suggested the existence of hydrogen-deficient post-AGB
stars \citep[e.g.\,][]{heap:75}. Later, the Palomar-Green Survey revealed a new spectral
class of H-deficient post-AGB stars, the PG1159 stars, which are dominated by
absorption lines of highly ionised He, C, and O \citep{wesemael:85}. It is now believed that the
PG1159 stars are descendants of the Wolf-Rayet CSPN and that the majority of
them will evolve into WDs with helium atmospheres (i.e. non-DA white dwarfs).

The origin of the hydrogen-deficiency in [WC] and PG1159 stars is
probably a late helium-shell flash, which means that a post-AGB star
(or WD) re-ignites helium-shell burning and transforms the star back
into an AGB star. This ``born-again AGB star'' phenomenon has been
discovered in early stellar evolution modeling
\citep{fujimoto:77,schoenberner:79} and was later invoked to explain
the H-deficiency observed in some hot post-AGB stars
\citep{iben:83a,herwig:99c}. A complete evolutionary track of such a
born-again evolution is shown as a dashed line in \abb{fig:hrd}. The
late He-shell flash causes complete envelope mixing. Hydrogen is
ingested and burnt and the surface chemistry of the star becomes
dominated by the composition of the previous He/C/O-rich intershell
layer (the region between the H- and He-burning shells of the AGB
star, see \kap{sec:agbevolution}).

In ``usual'', i.e. hydrogen-rich (post-) AGB stars intershell matter can be
dredged up to the surface (so-called third dredge-up) and results in the pollution
of the photosphere with intershell matter, that is, helium, 3$\alpha$-burning
products (C, O, Ne) as well as s-process elements which are produced by neutron
capture in the He-burning environment. The study of s-process element
abundances in AGB stars was and is the most important tool to reveal details of
the physics of burning and mixing processes. It is essential to understand these
processes, because they affect the yields from AGB stars which largely drive the
chemical evolution of the Galaxy.

In contrast to these H-rich stars, [WC] and PG1159 stars do not exhibit just
traces of intershell matter in their photospheres, but they are essentially made
up of intershell matter. This is because the mass of the intershell is much
larger than that of the hydrogen envelope and, thus, its composition dominates
the mixture of both layers triggered by the late He-flash. These H-deficient
post-AGB stars therefore offer the unique possibility to study intershell
abundances \emph{directly} and, hence, understand the physical processes leading
to the composition. 

\begin{deluxetable}{lrcccccccccccll}
\tabletypesize{\scriptsize} \rotate \tablecaption{Element abundances (mass
fractions) in representative hydrogen-deficient post-AGB stars of different
spectral classes. We discuss two possible evolutionary sequences: 1. [WCL]
$\longrightarrow$ [WCE] $\longrightarrow$ [WC]--PG1159 $\longrightarrow$ PG1159,
and 2. RCB $\longrightarrow$ extreme He-B-stars $\longrightarrow$ He-sdO
$\longrightarrow$ O(He)\label{tababu} } \tablewidth{0pt} \tablehead{
\colhead{\bf Spectral Class}&\colhead{\Teff}&\colhead{\logg}&&&&&&&&&&&&\\
\colhead{Star}          &\colhead{[K]}  &\colhead{[cgs]}& \colhead{H}&
\colhead{He}&  \colhead{C} &\colhead{N}& \colhead{O}& \colhead{F}&\colhead{Ne}&
\colhead{Si} & \colhead{S} & \colhead{Fe}& note& ref } \startdata
\multicolumn{9}{l}{\bf{[WCL]}}\\ 
IRAS\,21282   & 28\,000&3.2&.10   &.43   &.46    & $<$.005       &.01   &              &   &$<$.001&&    & H present         &2\\
PM1-188       & 35\,000&3.7&.01   &.39   &.47    &.01            &.07&&.03&.025&   &    & N present, Si high&2\\ 
He2-459       &77\,000&4.4&$<$.02&.40   &.50    &               &.10   &
&   &
&   &    & typical He/C/O    &2\\ \noalign{\smallskip}
\multicolumn{9}{l}{\bf{[WCE]}}\\ 
NGC\,1501     &134\,000&6.0&      &.50   &.35&               &.15   &              &   &    &   &    & typical He/C/O    &1\\
Sanduleak 3   &140\,000&6.0&      &.62   &.26    & .005          &.12   &&   &    &   &    & N present         &3\\ \noalign{\smallskip}
\multicolumn{9}{l}{\bf{[WC]--PG1159}}\\ 
Abell 78      &115\,000&5.5&      &.33& .50   & .02           & .15&1.0\,10$^{-5}$&   &    &&$<$.0001&Fe-deficient   &4,10,18\\ 
\noalign{\smallskip}\multicolumn{9}{l}{\bf{PG1159}}\\ 
HS1517+7403   &110\,000&7.0&      &.85   & .13&$<$3\,10$^{-5}$& .02  &              &   &    &   &    & C,O low           &5\\
HS2324+3944   &130\,000&6.2& .17  &.35   & .42   &$<$.0003\tablenotemark{a} &.06  & &   &    &   &    & H present         &11\\ 
PG1159-035&140\,000&7.0&$<$.02&.33   & .48   & .001         & .17  &3.2\,10$^{-6}$&.02&.00036&.0001&$<$.0003&typical He/C/O&6,7,9,10,19\\ 
PG1144+005    &150\,000&6.5&      &.38& .57   & .015          & .016 &1.0\,10$^{-5}$&.02&    &   &    & O low&8,9,10\\ 
\noalign{\smallskip} \tableline \noalign{\smallskip}
\multicolumn{9}{l}{\bf{RCB}}\\ 
RY Sgr        &7\,250&0.7&6\,10$^{-6}$&.98&.007& .003          & .0009&              &&.00040&.00045&.00020&            &12\\ 
\noalign{\smallskip}
\multicolumn{9}{l}{\bf{extreme He-B-stars}}\\
BD+10$^\circ$2179&16\,900&2.5&1\,10$^{-4}$&.98&.02& .0008        & .0004&&   &.00013&.00007&.00063&            &13\\ 
\noalign{\smallskip}
\multicolumn{9}{l}{\bf{He-sdO}}\\  
BD+37$^\circ$442&60\,000&4.0&$<$.001&.97&.025   & .003         &      &              &   &.00079& &    &&14\\ 
KS292         &  75\,000&5.5& .32  & .65  &.023   & .013         &      &&   &    &   &    &                   &15\\ 
\noalign{\smallskip}
\multicolumn{9}{l}{\bf{O(He)}}\\ 
K1-27         &105\,000&6.5&$<$.05&.98&$<$.015& .017          &      &              &   &    &   &    & N   present
&16\\ LoTr4         &120\,000&5.5& .11  &.89   &$<$.010& .003          &$<$.03&&   &    &   &    & H,N present       &16\\ 
HS1522+6615   &140\,000&5.5& .02&.97   & .01   &               &      &&   &    &   &    & H,C present       &16\\ 
\noalign{\smallskip} \tableline \noalign{\smallskip} 
{\bf Sun}     && & .73 &.25   &.0029  & .00089        &.0079&5.0\,10$^{-7}$&.0018&.00072&.00050&.0013&&17 \enddata
\tablenotetext{a}{uncertain} \tablerefs{ (1) \citet{koesterke:97b}; (2)
\citet{leuenhagen:98}; (3) \citet{koesterke:97}; 
(4) \citet{werner:92}; 
(5) \citet{dreizler:98a}; 
(6) \citet{werner:91}; 
(7) \citet{werner:96};  
(8) \citet{werner:91b}; 
(9) \citet{werner:04}; 
(10) \citet{werner:05b};
(11) \citet{dreizler:98b}; 
(12) \citet{asplund:00}; 
(13) \citet{pandey:05}; 
(14) \citet{bauer:95}; 
(15) \citet{rauch:91}; 
(16) \citet{rauch:98}; 
(17) \citet{grevesse:01}; 
(18) \citet{werner:03}; 
(19) \citet{jahn:05} }
\end{deluxetable}

\section{Quantitative spectral analyses}\label{observation}

In this section we will summarize results from spectral analyses. We will focus
on PG1159 stars (Sect.~\ref{sectpg1159}) because we think that in particular the
abundance analyses for many species is most advanced for this spectral
subclass. We will then turn to the [WC] central stars (Sect.~\ref{sectwc}) with
emphasis on evidence for the evolutionary link with the PG1159 stars. Both, [WC]
and PG1159 stars are thought to result from a late He-shell flash which,
however, is still debated. We will later discuss observations that appear to
challenge the ``born-again'' scenario and we will touch upon alternative
ideas. In any case, these stars are obviously exhibiting intershell matter and
thus are of immediate interest for insight into AGB evolution.

Besides [WC] and PG1159 stars, there exist several other spectroscopic
subclasses of H-deficient post-AGB stars which certainly cannot be explained by
the ``born-again'' scenario. These subtypes comprise RCB stars, extreme helium
B-stars, helium-rich subdwarf O stars, and the O(He) central stars. They all
generally have helium-dominated atmospheres in contrast to [WC] and PG1159 stars
which usually have high carbon abundances. These helium-dominated subtypes might
form a distinct post-AGB evolutionary channel being caused by a stellar merging
event. Because the origin of their abundance patterns is unclear, they are -- at
the moment -- not useful for probing AGB evolution. Therefore we will discuss
these subtypes for completeness only (Sect.~\ref{sectrcrb}), to order the zoo of
H-deficient post-AGB stars.

Table~\ref{tababu} gives an overview about typical abundance patterns found
among all the discussed spectral subclasses. The [WC] stars are sub-divided into
late-type and early-type stars, [WCL] and [WCE]. In addition, a [WC]-PG1159
transition type has been introduced, denoting objects with a mixed
emission/absorption line spectrum.

\begin{deluxetable}{rccllcccll}
\tabletypesize{\scriptsize} \tablecaption{ The known PG1159 stars and results of
spectroscopic analyses. It is also noted if the star is a pulsating variable
object and if it has a planetary nebula. Objects with detectable residual
hydrogen are denoted ``hybrid'' stars.
\label{tabpg1159}
} \tablewidth{0pt} \tablehead{ Star                       &\Teff   &\logg&
C/He\tablenotemark{a}& O/He &Mass& Var.&PN&
ref.\tablenotemark{b}&remark\tablenotemark{c}\\ &[1000 K]&[cgs]&
&      &[M$_\odot$]&&& } \startdata
%                                        C/He   O/He   mass   var   PN   ref
H\,1504+65                 & 200 & 8.0 &$>$50 &$>$50 & 0.89 & no  & no  & 1  &
He-deficient\\   RX\,J0122.9$-$7521         & 180 & 7.5 & 0.3  & 0.17 & 0.72 &
no  & no  & 2  & \\  RX\,J2117.1+3412           & 170 & 6.0 & 1.4  & 0.16 & 0.70
& yes & yes & 3  & \\  HE\,1429$-$1209            & 160 & 6.0 & 1.4  & 0.16 &
0.67 & yes & no  & 4  & \\ PG\,1520+525               & 150 & 7.5 & 0.9  & 0.4
& 0.67 & no  & yes & 3  & \\  PG\,1144+005               & 150 & 6.5 & 1.5  &
0.05 & 0.60 & no  & no  & 3  & \\  Jn\,1                      & 150 & 6.5 & 1.5
& 0.8  & 0.60 & no  & yes & 5  & \\  NGC\,246                   & 150 & 5.7 &
0.5  & 0.1  & 0.72 & yes & yes & 3  & \\  PG\,1159$-$035             & 140 & 7.0
& 1.5  & 0.5  & 0.60 & yes & no  & 3  & \\  NGC\,650                   & 140 &
7.0 & ?    & ?    & 0.60 & no  & yes & 6  & :\\  Abell\,21=Ym29             &
140 & 6.5 & 1.5  & 0.5  & 0.58 & no  & yes & 2  & :\\  K\,1$-$16
& 140 & 6.4 & 1.5  & 0.5  & 0.58 & yes & yes & 3  & :\\  Longmore 3
& 140 & 6.3 & 1.4  & 0.16 & 0.59 & no  & yes & 2  & :\\  PG\,1151$-$029
& 140 & 6.0 & 1.5  & 0.5  & 0.60 & no  & no  & 2  & :\\  VV\,47
& 130 & 7.0 & 1.5  & 0.4  & 0.59 & no  & yes & 5  & :\\  HS\,2324+3944
& 130 & 6.2 & 1.2  & 0.16 & 0.58 & yes & no  & 11 & : hybrid \\
SDSS\,J102327.41$+$535258.7& 120 & 7.0 & 0.9  & ?    & 0.58 & no  & no  & 7  &
\\ Longmore 4                 & 120 & 5.5 & 0.9  & 0.2  & 0.65 & yes & yes & 3
& \\  SDSS\,J001651.42$-$011329.3& 120 & 5.5 & 0.6  & 0.16 & 0.65 &     & no  &
7  & \\ PG\,1424+535               & 110 & 7.0 & 0.9  & 0.12 & 0.57 & no  & no
& 3  & \\  HS\,1517+7403              & 110 & 7.0 & 0.15 & 0.02 & 0.57 & no  &
no  & 9  & \\  Abell\,43                  & 110 & 5.7 & 1.2  & 0.16 & 0.59 & yes
& yes & 11 & : hybrid \\ NGC\,7094                  & 110 & 5.7 & 1.2  & 0.16 &
0.59 & no  & yes & 11 & : hybrid \\ SDSS\,J075540.94+400918.0  & 100 & 7.6 &
0.09 & ?    & 0.62 &     & no  & 7  & \\ HS\,0444+0453              & 100 & 7.5
& ?    & ?    & 0.59 &     & no  & 8  & :\\  IW\,1                      & 100 &
7.0 & ?    & ?    & 0.56 & no  & yes & 6  & :\\  Sh\,2$-$68                 &
96 & 6.8 & ?    & ?    & 0.55 &     & yes & 10 & : hybrid \\
SDSS\,J144734.12+572053.1  &  95 & 7.6 & 0.1  & ?    & 0.61 & no  & no  & 7  &
\\ PG\,2131+066               &  95 & 7.5 & 0.9  & 0.4  & 0.58 & yes & no  & 9
& \\  SDSS\,J134341.88+670154.5  &  90 & 7.6 & 0.15 & ?    & 0.60 & no  & no  &
7  & \\ MCT\,0130$-$1937           &  90 & 7.5 & 0.3  & 0.04 & 0.60 & no  & no
& 2  & \\  PG\,1707+427               &  85 & 7.5 & 0.9  & 0.4  & 0.59 & yes &
no  & 3  & \\  PG\,0122+200               &  80 & 7.5 & 0.9  & 0.4  & 0.58 & yes
& no  & 9  & \\  HS\,0704+6153              &  75 & 7.0 & 0.3  & 0.12 & 0.51 &
& no  & 9  & \\  NGC\,6852                  &     &     &      &      &      &
& yes & 6  & K1$-$16 type\\ NGC\,6765                  &     &     &      &
&      &     & yes & 6  & K1$-$16 type\\ Sh\,2$-$78                 &     &
&      &      &      &     & yes & 6  & PG\,1424+535 type    \enddata
\tablenotetext{a}{Abundance ratios by mass} \tablenotetext{b}{Reference to the
most recent spectroscopic work} \tablenotetext{c}{Colons denote uncertain,
preliminary results of spectroscopic analyses} \tablecomments{No spectroscopic
analysis of the last three stars was performed.  [WC]-PG1159 transition type
objects like Abell~78 and Abell~30 are not listed.}  
\tablerefs{ (1) \citet{werner:04b}; 
(2) \citet{werner:04}; 
(3) \citet{werner:05b}; 
(4) \citet{werner:04c};  
(5) \citet{rauch:95}; 
(6) \citet{napiwotzki:95}; 
(7) \citet{huegelmeyer:05}; 
(8) \citet{dreizler:95};  
(9) \citet{dreizler:98a};
(10) \citet{napiwotzki:99b}; 
(11) \citet{dreizler:98b}
}
\end{deluxetable}

\subsection{PG1159 stars}\label{sectpg1159}

\subsubsection{Spectral classification}

The optical spectra of PG1159 stars are characterized by weak and broad
absorption lines of \ion{He}{2} and \ion{C}{4}, sometimes with central emission
reversals. The hottest objects also display \ion{O}{6} and \ion{Ne}{7}
lines. Three spectral subclasses have been introduced which allow a coarse
characterisation of each star. According to the appearance of particular line
features the subtypes ``A'' (absorption lines), ``E'' (emission lines), and lgE
(low gravity; emission lines) were defined \citep{werner:92b}.

\subsubsection{General characteristics: Temperature, gravity and He/C/O
abundances; mass-loss rates}

First quantitative spectral analyses became feasible with the construction of
line blanketed non-LTE model atmospheres accounting for the peculiar chemical
composition \citep{werner:91}. At that time, only a handful of PG1159 stars
was identified. Today, 37 PG1159 stars are known. Most of them were found by
systematic spectroscopic observations of central stars of old, evolved planetary
nebulae 
\citep{napiwotzki:95} as well as follow-up spectroscopy of
faint blue stars from various optical sky surveys (Palomar-Green Survey,
Montreal-Cambridge-Tololo Survey, Hamburg-Schmidt and Hamburg-ESO Surveys, Sloan
Digital Sky Survey) and soft X-ray sources detected in the ROSAT All Sky
Survey. In Table\,\ref{tabpg1159} we give a complete list of all known PG1159
stars and their location in the \Teff--\logg diagram is shown in
Fig.\,\ref{fighrd}.  It is seen that they span a wide temperature and gravity
range. They represent stars in their hottest phase of post-AGB evolution. Some
of them (those with \logg $\lappr$ 6.5) are still helium-shell burners (located
before the ``knee'' in their evolutionary track) while the majority has already
entered the WD cooling sequence.

Estimates for mass and luminosity can be obtained by comparison with theoretical
evolutionary tracks. The mean mass of PG1159 stars is 0.61~M$_\odot$ using the
older models of H-rich central stars (solid and dashed lines in
\abb{fighrd}). However, the H-deficient $0.604\msun$ track (dashed-dotted lines
in \abb{fighrd}) is systematically hotter than the older tracks, and indicates
that the  central stars are on average in fact somewhat less massive than
that. It is important to note, that the higher temperature of the new tracks is
not the result of being H-deficient. Instead, the reason is the more realistic
AGB progenitor evolution that now includes the third dredge-up after most of
the AGB thermal pulses. This leads to a higher luminosity for a given core mass
and during the post-AGB evolution to higher temperature (\kap{sec:hrdnew}).

Optical spectra are dominated by numerous \ion{He}{2} and \ion{C}{4} lines from
which the He/C ratio can be derived. Only the hottest objects (\Teff $\gappr$
120\,000\,K) exhibit oxygen lines, too, (\ion{O}{6} and, sometimes, very weak
\ion{O}{5}) so that the O abundance in cooler stars cannot be determined unless
UV spectra are available. Similarly, only the hottest objects display optical
neon lines (\ion{Ne}{7}) and only UV spectra allow access to the neon abundance
in the case of cooler objects. Hydrogen poses a special problem, because all H
lines are blended with \ion{He}{2} lines. In medium-resolution ($\approx$1~\AA)
optical spectra hydrogen is only detectable if its abundance is higher than
about 0.1 (all abundances in this paper are given in mass fractions). With
high-resolution spectra, which are difficult to obtain because of the faintness
of most objects, this limit can be pushed down to about 0.02. For PG1159 stars
within a PN the situation is even more difficult because of nebular Balmer
emission lines.

\begin{figure}
\includegraphics[width=\columnwidth]{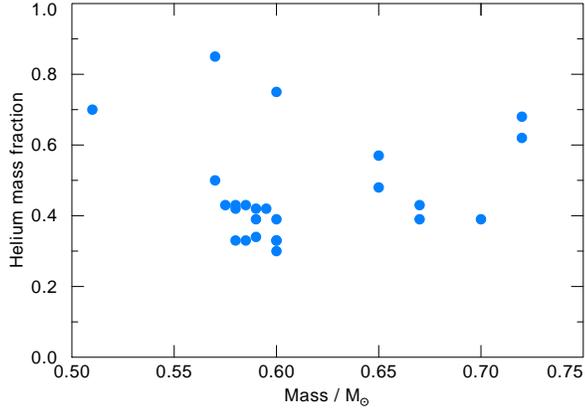}
\caption{Helium mass fraction in PG1159 stars as a function of stellar mass.}\label{fighelium}
\end{figure}

\begin{figure}
\includegraphics[width=\columnwidth]{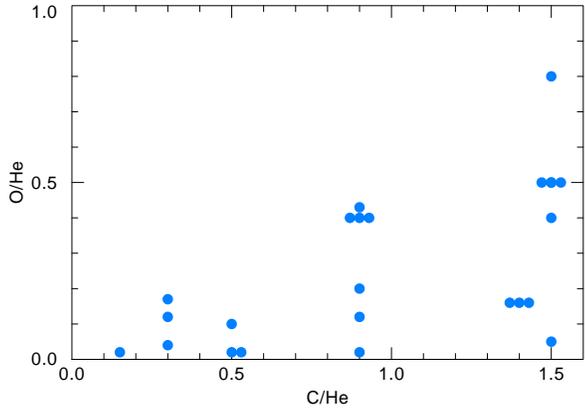}
\caption{Carbon and oxygen abundance ratios relative to helium (by mass) in PG1159 stars.}\label{figabu}
\end{figure}

The He, C, and O abundances show strong variations from star to star, however, a word of
caution is appropriate here, too. The quality of the abundance determination is
also very different from star to star. For some objects only relatively poor
optical spectra were analysed, while others were scrutinized with great care
using high S/N high-resolution optical {\em and} UV/FUV data. Nevertheless, we
think that the abundance scatter is real. The prototype PG1159-035 displays what
could be called a ``mean'' abundance pattern: He/C/O=0.33/0.50/0.17. An extreme
case, for instance, with low C and O abundances is HS1517+7403, having
He/C/O=0.85/0.13/0.02. Taking all analyses into account, the range of mass
fractions for these elements is, approximately: He=0.30--0.85, C=0.15--0.60,
O=0.02--0.20 (excluding the peculiar object H1504+65, see
Sect.~\ref{secth1504}).  There is a strong preference for a helium abundance in
the range 0.3--0.5,  independent of the stellar mass  (Fig.~\ref{fighelium}).
Only a minority of stars has a higher He abundance, namely in the range
0.6--0.8.  There is a tendency that a high O abundance is only found in objects
with a high C abundance (Fig.~\ref{figabu}).

Some remarks on the possible analysis errors are necessary. As just pointed out,
the observational data are of rather diverse quality, but in general the
following estimates hold. The temperature determination is accurate to
10--15\%. The surface gravity is uncertain within 0.5~dex. Element abundances
should be accurate within a factor of two. The main problem arises from
uncertainties in line broadening theory which directly affects the gravity
determination and the abundance analysis of He, C, and O.

\subsubsection{Objects with residual hydrogen: Hybrid PG1159 stars}

We already mentioned that hydrogen is difficult to detect. However, four objects
clearly show Balmer lines and they are called hybrid PG1159 stars. The deduced H
abundance is quite high: H=0.17 \citep[Tab.~\ref{tababu}, ][]{dreizler:98b}. It is
worthwhile to note for the discussion on their evolution, that nitrogen is seen
in the optical spectra of some of these stars but quantitative analyses are
still lacking. The hybrid star NGC\,7094 shows Ne and F enhancements like many
PG1159 stars (see below). So one can conclude that, aside from the presence of
H, the element abundance pattern of the hybrid PG1159 stars seems to resemble
that of many other PG1159 stars. However, all the hybrids have not yet been
analysed appropriately although good UV and optical spectra are available.

\subsubsection{Nitrogen and pulsation instability}

Some PG1159 stars do show nitrogen lines while others, with similar
temperature and gravity, do not. These are \ion{N}{5} lines in the
optical wavelength range and the resonance line doublet in the UV at
1239/1243~\AA. The derived N abundance is of the order 0.01
\citep{werner:91b,dreizler:98a}. The presence of N shows no clear
correlation to the relative abundances of the main atmospheric
constituents (He, C, O). But a remarkable correlation between the
presence of N and the pulsational instability of PG1159 stars has been
found: All four pulsating objects in a sample of nine examined PG1159
stars are showing nitrogen \citep{dreizler:98a}. Among these
objects is the prototype PG1159-035, for which we have recently taken
a high-resolution HST/STIS spectrum, allowing to separate the
interstellar and photospheric components of the \ion{N}{5} resonance
doublet. As a consequence, we find that the N abundance is distinctly
lower, namely 0.001 \citep[][ and Reiff \etal in prep.]{jahn:05,reiff:05,reiff:06a}, which weakens
the nitrogen/pulsation correlation. In any case, such a correlation is
difficult to explain with current stellar pulsation models 
\citep[e.g. ][]{quirion:04}. 
Due to its low abundance, nitrogen cannot affect
pulsational properties, but it was speculated that N is a marker for
different evolutionary histories leading to different He/C/O
abundances in the pulsation driving regions. More about PG1159 pulsators follows
in Sect.~\ref{sectseismology}.

\subsubsection{Neon}

Neon has been detected in ten PG1159 stars by the identification of \ion{Ne}{7}
lines in optical and FUV spectra 
\citep{werner:94,werner:04}. The
derived abundances are of the order 0.02, i.e. roughly ten times solar. Only the
hottest and most luminous objects are able to ionize neon strongly enough to
show these lines. Unfortunately, no lines from lower ionization stages are
observed, so that the neon abundance in cooler PG1159 stars remains unknown. The
strongest \ion{Ne}{7} line is found in FUSE spectra, located at 973\AA. The
absorption line core almost reaches zero intensity in some objects. In the most
luminous objects (e.g. K1$-$16) this line displays a powerful P~Cygni profile
\citep{herald:05}.

\subsubsection{Fluorine}

Lines from \ion{F}{5} and \ion{F}{6} have been discovered in FUSE spectra of
several PG1159 stars \citep{werner:05b}. The derived abundances show a
surprisingly strong variation from star to star, ranging between solar and 250
times solar. Correlations between the F and other element abundances are not
obvious. We have speculated that the variety of F abundances is a consequence of
the stellar mass, which according to \citet{lugaro:04a} strongly affects the
production of fluorine.

\subsubsection{Other light metals: Si, P, S}\label{Si}

Lines from silicon, phosphorus, and sulfur were discovered in FUSE spectra of
several PG1159 stars. Analyses are currently performed and we can report
preliminary results only \citep[][ and Reiff \etal in prep.]{jahn:05,reiff:05,reiff:06a}.

The silicon abundance in PG1159 stars is of interest, because a strong
overabundance (up to Si=0.03, i.e. 40 times solar) has been found in a handful of
[WCL] stars \citep{leuenhagen:98}. The photospheric component of the
\ion{Si}{4}~1394/1402~\AA\ resonance doublet has been identified in the above
mentioned HST/STIS high resolution spectrum of the prototype. Our best line fit
is obtained with Si=0.5~solar, which means no significant deviation from the
solar abundance within error limits. Another silicon line pair,
\ion{Si}{4}~1122/1128~\AA, can be identified in relatively cool objects
only. For PG1424+535 we find a solar silicon abundance from these
lines and an upper limit of 0.1 solar for PG1520+525.

Phosphorus was discovered in a number of PG1159 stars by the identification of
the \ion{P}{5} resonance doublet at 1118/1128~\AA. First results for two objects
indicate a roughly solar abundance.

The most prominent sulfur feature is the \ion{S}{6}~933/945~\AA\ resonance
doublet. It was first detected in K1-16 and a solar S abundance was derived
\citep{miksa:02}. It is also visible in other PG1159 stars. In the prototype 
star and two other objects we find a surprisingly strong S depletion (less than
0.1 solar).

\subsubsection{Iron deficiency}

Model spectra for PG1159 stars predict detectable lines from
\ion{Fe}{6} and \ion{Fe}{7} in the UV for objects which are not too
hot (\Teff$\leq$140\,000~K), otherwise iron is ionised to even higher
stages which have lines only in the EUV range beyond the Lyman
edge. But iron lines are narrow and require high-resolution
($\approx$0.1~\AA) high-S/N spectra for a quantitative analysis. Up to
now, iron has not been detected in \emph{any} PG1159 star. In most
cases, a solar Fe abundance cannot be excluded, because the data
quality does not allow more stringent conclusions. For two PG1159
stars, however, it was shown that the lack of iron lines means that
the iron abundance is subsolar by at least one dex \citep[K1-16 and
NGC~7094; ][]{miksa:02}. A similar result was obtained for the
[WC]-PG1159 transition object Abell~78 \citep{werner:03} and several
[WC] stars (see below). The prototype PG1159-035 is subsolar in Fe by
at least a factor of five \citep{jahn:05}. The suggestion of
\citet{2003IAUS..209...85H} that iron has been transformed into
heavier elements will be discussed below. Our search for lines from
elements heavier than Fe has so far been unsuccessful.

\subsubsection{Unidentified lines}

About two dozen of absorption lines in the 1000--1700~\AA\ region of PG1159
stellar spectra taken with FUSE, HST, and IUE remain unidentified. Most of them
are commonly seen in at least two objects, and many are quite prominent. For
example, one of the strongest unidentified absorption lines in the UV spectrum of
the prototype as well as of other PG1159 stars is located at 1270.2~\AA. It
always appears together with a small number of weaker absorption lines in the
1264--1270~\AA\ range. It could be a hitherto unidentified neon or magnesium
multiplet.

Also, a number of weak lines in high-quality optical spectra remain
unidentified, too.  We have recently identified one of these features as a
\ion{Ne}{7} multiplet, whose wavelength was only inaccurately known before \citep{werner:04}.

We are confident that all the unidentified lines do not stem from C or O. It is
obvious on the other hand that line lists from other highly ionized metals are
rather incomplete, inaccurate or even non-existent.  Some of the unidentified
lines might stem from other light metals (Mg, Al) or even from heavy metals
beyond the iron group, e.g., s-process enhanced elements. We should make every
effort to identify these lines. New species could be found and their abundances
checked against evolutionary models.

\subsubsection{Mass loss; occurrence of ultra-high ionisation lines}

PG1159 stars do not exhibit wind features in their optical spectra. UV
spectroscopy, however, reveals that many of the low-gravity central
stars display strong P~Cygni profiles, e.g. of the resonance lines
from \ion{C}{4} and \ion{O}{6}; and of subordinate lines from
\ion{He}{2}, \ion{O}{5}, and \ion{Ne}{7}. Mass-loss rates in the range
\md $=-8.3\,...\,-6.9$ were derived
\citep{koesterke:98b,koesterke:98,herald:05} and it appears that they
are in accordance with predictions from radiation driven wind theory.

The central star of Longmore~4 showed a remarkable event, turning its
spectral type from PG1159 to [WCE] and back again to PG1159, most
probably due to a transient but significant increase of the mass-loss
rate \citep{werner:92c} The reason is unknown but might be connected
to the fact that the star is a pulsator. This phenomenon has never
been witnessed again, neither in Longmore~4 nor any other PG1159 star.
The outburst phenomenon observed in the [WR] central star of the LMC
planetary nebula N66 is probably the consequence of mass-transfer
within a close-binary system (see Sect.~\ref{n66}).

The most luminous PG1159 stars display ultrahigh-ionisation
\emph{emission} lines, the most prominent one is \ion{O}{8}~6068~\AA\
\citep[e.g. ][]{werner:94b}. This phenomenon is also seen in [WCE]
stars and in [WC]--PG1159 transition objects, as well as in the
hottest known DO white dwarf (KPD0005+5106).  It is clear that the
photospheric temperatures are not high enough to produce these lines
and it is possible that they arise from shock-heated regions in the
stellar wind.

A large fraction of all DO white dwarfs shows such ultrahigh-ionisation lines in
\emph{absorption} \citep[e.g. \ion{C}{6}, \ion{N}{7}, \ion{O}{8},
\ion{Ne}{10},][]{werner:95}. 
This remarkable and still unexplained phenomenon was recently
discovered for the first time in a new PG1159 star \citep{huegelmeyer:06}. In
addition, the usual photospheric absorption lines (\ion{He}{2}, \ion{C}{4}) are
much stronger than predicted from any model and this behavior, too, is shared
with that of the respective DO white dwarfs.

We remark that, even more strange, too-deep photospheric \ion{He}{2}
absorption lines are exhibited by some DO white dwarfs which do
\emph{not} show ultrahigh-ionisation absorption lines at the same time
\citep[e.g. ][]{werner:04c}. A respective PG1159-type counterpart has
also been discovered \citep{nagel:06}.

\subsubsection{Peculiar: H1504+65}\label{secth1504}

H1504+65 displays an optical spectrum that is at first sight rather similar to
other hot high-gravity PG1159 stars. However, it has been shown that this
object is helium-deficient and its atmosphere is mainly composed of C and
O. Probably it has an evolutionary history that is different from all the other
PG1159 stars, hence, we will not further discuss this interesting peculiar
object \citep[see ][ for details]{werner:04b}.

\subsubsection{Descendants: DA and non-DA white dwarfs}

The coolest high-gravity PG1159 star has \Teff=75\,000~K and the hottest white
dwarfs have \Teff$\approx$120\,000~K. In this temperature range the PG1159 stars
must be turned into a DA or non-DA white dwarf by gravitational settling,
depending on the presence of residual hydrogen in the PG1159 stellar
envelope. It is tempting to assume that the majority of the PG1159 stars will
become DO and then DB and DQ white dwarfs, but it is unknown to which extent this is
true. Clearly, the hybrid-PG1159 stars will become DA white dwarfs, but this
could in principle also hold for all other PG1159 stars because a H abundance as
high as 0.01 cannot be excluded by spectroscopic means. Depending on the amount
of residual H in the PG1159 envelope, DAs with different hydrogen layer mass
will emerge. It can be speculated that the ZZ~Ceti stars (DA pulsators) for
which thin H envelopes were inferred are descendants of PG1159 stars \citep{althaus:05b}.

\begin{deluxetable}{rccllcccll}
\tabletypesize{\scriptsize} \tablecaption{Results from asteroseismology of
PG1159 stars and the [WC4] central star of NGC~1501. We compare the stellar mass
derived by spectroscopic means M$_{\rm spec}$ (from Tab.~\ref{tabpg1159}) with
the pulsational mass M$_{\rm puls}$. Other columns list envelope mass M$_{\rm
env}$ (all masses in solar units) and rotation period P$_{\rm rot}$ in days.
\label{tabpulsators}
} \tablewidth{0pt} \tablehead{ Star            &M$_{\rm spec}$&M$_{\rm
puls}$&M$_{\rm env}$&P$_{\rm rot}$&ref  } \startdata 
PG\,2131+066    & 0.58         & 0.61         & 0.006       & 0.21 & 1 \\ 
PG\,0122+200    & 0.58         & 0.59         &             & 1.66 & 2 \\ 
RX\,J2117.1+3412& 0.70         & 0.56         & 0.045       & 1.16 & 3 \\ 
PG\,1159$-$035  & 0.60         & 0.59         & 0.004       & 1.38 & 4 \\ 
PG\,1707+427    & 0.59         & 0.57         &             &      & 5 \\ 
NGC 1501        &              & 0.55         &             & 1.17 & 6
\enddata \tablerefs{
(1) \citet{kawaler:95}; 
(2) \citet{fu:06}; 
(3) \citet{vauclair:02a}; 
(4) \citet{kawaler:94}; 
(5) \citet{kawaler:04};
(6) \citet{bond:96}
}
\end{deluxetable}

\subsubsection{Results from asteroseismology}\label{sectseismology}

\begin{figure}
\includegraphics[width=\columnwidth]{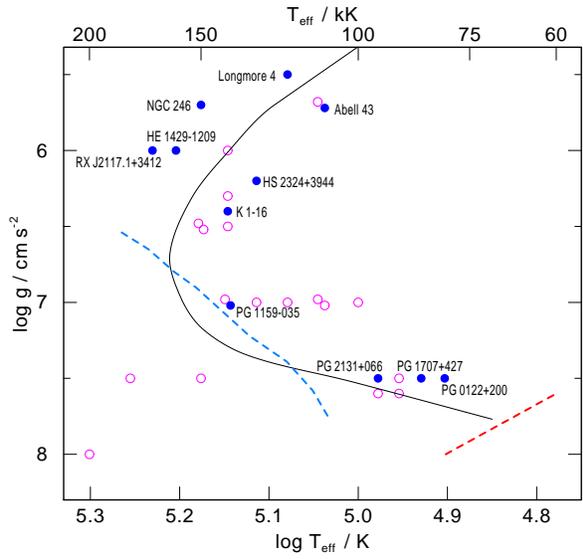}
\caption{The pulsating PG1159 stars (filled symbols with name-tags)
and the non-pulsators. The dashed lines are theoretical blue and red
edges of the instability strip from \citet{gautschy:05} and
\citet{quirion:04}, respectively. See text for discussion about these
edges and the co-existence of non-pulsators in the strip
(Sect.~\ref{sectseismology}). Also shown is the 0.6~M$_\odot$ post-AGB
track from Fig.~\ref{fighrd}.}\label{figpulsators}
\end{figure}

Some of the PG1159 stars are non-radial g-mode pulsators (see
Tab.~\ref{tabpg1159}) and they define the GW~Vir (=PG1159-035) instability strip
in the HRD (Fig.~\ref{figpulsators}). These variables are among the best studied
by asteroseismologic analyses. Such investigations on the interior of PG1159
stars hold important clues for their origin.

In Tab.~\ref{tabpulsators} we summarize the results from asteroseismology of five
PG1159 stars plus one [WCE] central star. The agreement between pulsational and
spectroscopic mass determinations is very good (within 5\%) except for
RX\,J2117.1+3412, where the difference is of the order 20\%. This is significant
because in order to shift its position in
the \logg--\Teff\ diagram onto the M$_{\rm puls}$=0.56~M$_\odot$ track, \Teff\
would need to be decreased from 170\,000~K to $<$120\,000~K, which is clearly
ruled out by detailed optical and UV/FUV spectroscopy. The mass discrepancy is
possibly due to inadequate pulsation models, a weakness in their asteroseismic
analyses pointed out by \citet{vauclair:02a}.

Of considerable interest for comparison with evolutionary calculations is the
analysis of the mode-trapping features in the period spacings which allows to
investigate the stellar interior structure. Tab.~\ref{tabpulsators} lists the
results obtained for the stellar envelope mass M$_{\rm env}$, i.e.\ the mass
above the chemical discontinuity between the He-rich envelope and the C/O
core. These envelope masses are much smaller than the intershell region 
remaining on top of the C/O core (M=$2\cdot10^{-2}$~M$_\odot$) immediately after
a (very) late thermal pulse. Observational mass-loss determination for FG~Sge by \citet{gehrz:05} show
that born-again stars have high mass-loss rates over prolonged periods of time
(see Sect.~\ref{kap:hist}). These may remove enough mass to indeed identify the
envelope mass from pulsation analyses with the mass between the He-free
core and the surface.
 Note, however, that
recent investigations involving complete evolutionary models come to the
conclusion that the mode-trapping features could result from structures in the
C/O core \citep{corsico:05}.

Recent theoretical pulsation driving modeling could clarify important
problems. Early stability calculations \citep[e.g.][]{starrfield:83}
suggested that the C/O $\kappa$-mechanism in PG1159 stars is suffering
from the so-called H and He poisoning phenomenon. Accordingly, the He
abundance in the photospheres, and particularly the H abundance in the
hybrid-PG1159 stars, is too high to drive pulsations in the
sub-photospheric layers, hence, abundance gradients had to be
invoked. It is however difficult to explain how such an abundance
gradient could be maintained. The $\kappa$-mechanism operates at
depths where T$\approx 10^6$~K, which lies in the outer $\approx
10^{-8}$~M$_\star$ mass fraction of the star \citep[e.g.,
][]{quirion:04}. In the case of RX\,J2117.1+3412, for instance, the
observed mass-loss rate of \md $=-7.4$ \citep{koesterke:98b} implies
that the material in the driving region is appearing on the surface
and renewed on a very short time scale, namely three months.  New
pulsation modeling with improved opacity tables has shown that this
poisoning problem is not so severe and, hence, no abundance gradients need to be
invoked \citep{saio:96,gautschy:97,gautschy:05,quirion:04}.
They can even explain the presence of pulsations in the H-rich hybrid-PG1159 stars
\citep{quirion:05a}.  However, these results are still at odds with
similar calculations performed by others \citep{cox:03}.

The co-existence of pulsators and non-pulsators in the instability
strip can be explained in terms of differences in the chemical surface
composition. The red edge of the strip is essentially identical with
the location of the coolest high-gravity PG1159 stars. Gravitational
diffusion of C and O out of the remaining He-envelope removes the
driving agents and turns the PG1159 star into a DO white dwarf
\citep{quirion:05b}.  There is no sharp blue edge of the strip, as its
position is different for each star, depending on its chemical
composition. Asteroseismic analyses of DB white dwarfs confirm the diffusion
scenario and the PG1159--DO--DB evolutionary link \citep{metcalfe:05}.

\subsection{[WC] central stars}\label{sectwc}

[WC] central stars exhibit spectra that are very similar to their massive
counterparts. As a consequence of the higher mass-loss rates \citep[\md
$=-7.0\,...\,-4.9$, ][]{koesterke:01} compared to PG1159 stars, they show
essentially pure emission line spectra, mainly presenting broad and bright lines
of He, C, and O.

Recent summaries of [WC] identifications and spectral analyses can be found in
\citet{acker:03} and \citet{hamann:03a}, 
respectively.

\subsubsection{Abundances in [WC] stars compared to PG1159 stars}

The [WC] stars are divided into two subgroups, namely late-type and
early-type objects: [WCL] and [WCE]. From their location in the
\logg--\Teff\ diagram (Fig.~\ref{fighrd}) it is suggestive that they
form an evolutionary sequence, but this is at odds with the current
state of quantitative spectral analyses. The mean carbon abundance in
[WCL] stars is 0.50 while it is lower by about a factor of two in the
[WCE]s \citep{koesterke:01}.  In recent re-analyses with new models
this difference does not disappear \citep{hamann:05a}. It is
conceivable that systematic errors occur because the
abundance analyses rely on different spectral lines in [WCL]s and
[WCE]s.

Aside from this discrepancy it seems obvious that the PG1159 stars are
the progeny of [WC] stars, because they can accommodate the variety of
He/C abundance ratios observed in the [WC]s. To corroborate this
evolutionary link it is useful to compare other element abundances. We
refer to work already quoted in this section and to
\citet{leuenhagen:98}.

Hydrogen: In three [WCL] stars H was detected and the abundances range between
0.1 and 0.01, however, it has been questioned that the Balmer lines are of
photospheric origin \citep{demarco:01a}. No H was detected in [WCE] stars, but no strict upper limits can
be given (H$\lappr$0.1).  This is in line with upper limits of about 0.1 for the
H abundance in PG1159 stars and the H abundance of 0.17 in the hybrid-PG1159
stars.

Nitrogen: Some [WCL] stars exhibit N lines and abundances of the order 0.01 were
found. Also, some [WCE] show N lines and the derived abundances are just
slightly lower, namely 0.003--0.005.  These abundances are very similar to those
determined for objects in the PG1159 group.

Oxygen: The range of O abundances observed in [WC]s is very similar to that of
the PG1159 stars.

Neon: \ion{Ne}{1} lines were detected in four [WCL] stars and abundances of the
order 0.03 were derived. Again, this is very similar to the PG1159 group.

Silicon: As already mentioned above (Sect.~\ref{Si}), strong Si overabundances
in some [WCL] stars were found, which is at odds with the approximately solar Si abundance we
find in the PG1159 prototype. More analyses are needed to see if there exists a
large abundance scatter within the [WCE] or PG1159 stars.

Iron: Like in the case of the PG1159 stars, the Fe abundance in [WC]
stars has been examined in a few cases only. An iron deficiency has
been found, being in qualitative agreement with PG1159 stars.
\citet{graefener:03} report a low Fe abundance in SMP\,61, a [WCE]
star in the LMC. Its abundance is at least 0.7~dex below the LMC
metallicity. \citet{crowther:98} find evidence for an underabundance
of 0.3--0.7~dex in the Galactic [WCL] stars NGC~40 and BD+30$^\circ$3639.

\subsubsection{[WC]--PG1159 transition objects; weak emission line stars (WELS)}

The [WC]--PG1159 class comprises stars which show both, spectral characteristics
of PG1159 stars and [WC] stars, namely mixed absorption/emission line
spectra. Only two objects can be safely assigned to this group (Abell~30,
Abell~78). They have He/C/N/O abundances similar to PG1159 and [WC] stars
\citep{werner:92}.

It must be emphasized that the class of so-called ``weak emission line stars''
(WELS), which comprises dozens of central stars 
\citep[e.g. ][]{tylenda:93}, 
is \emph{not} identical with the [WC]--PG1159 class. The WELS are poorly studied
quantitatively, and some of them are clearly ``usual'' hydrogen-rich post-AGB
stars \citep{mendez:91,fogel:03}.

\subsubsection{[WO] and O\,VI classifications}

The term ``O\,VI sequence'' was coined by \citet{smith:69} to denote
central stars with the most highly excited optical stellar
spectra. The distinguishing feature of this group is the presence of
the \ion{O}{6}~3811/3834~\AA\ emission doublet. In today's
terminology, the O\,VI stars is a collective name for the hottest
[WCE] and PG1159 stars and the [WC]--PG1159 transition objects, which
all show \ion{O}{6} emission lines.

The WO class (O for oxygen) was first introduced by \citet{barlow:82}
for WR stars with highest excitation spectra.  It was also introduced
in refined classification systems that are applicable to both, massive
WR stars and WR central stars \citep{crowther:98,acker:03}.  Like in
the ``O\,VI sequence'' such spectra are exhibited by the hottest
objects showing \ion{O}{6} emission lines. In essence, the [WO]
classification (with subclasses [WO1] to [WO4]) is used alternatively
for the earliest [WCE] subtypes [WC2] and [WC3].

\subsubsection{Are there [WN] central stars?}\label{n66}

It is believed that there are two central stars that exhibit a spectrum similar
to massive WN stars, i.e., it is dominated by N instead of C emission lines. For
one of these objects (PM5) it cannot be completely ruled out that it is in fact a
massive WN star with a ring nebula \citep{morgan:03a}. The other object is the
LMC planetary nebula N66. It is thought that this object is a white dwarf
accreting matter from a close companion \citep{hamann:03b} and, hence, its
surface composition is not the result of single-star evolution.

\subsection{He-dominated post-AGB objects: RCB, extreme He-B, He-sdO, and
O(He) stars}\label{sectrcrb}

There exists a small group of four extremely hot objects
(\Teff$>$100\,000~K) which have almost pure \ion{He}{2} absorption
line spectra in the optical. As introduced by \citet{mendez:91}, these
stars are classified as O(He). Their atmospheres are indeed helium
dominated, only trace amounts of CNO elements are detected \citep[][ see
 also Tab.~\ref{tababu} for atmospheric parameters]{rauch:98}. In
the \logg--\Teff\ diagram they are found among the PG1159 stars.
While the born-again evolutionary models can explain the rich
diversity of different He/C/O patterns in [WC] and PG1159 stars, they
never result in such helium-dominated surface abundances. It is
therefore natural to speculate on the existence of a third post-AGB
evolutionary sequence and its origin. The O(He) stars could be the
long-searched progeny of the RCB stars, which are relatively cool
(\Teff\ $<$ 10\,000\,K) stars with helium-dominated atmospheres,
too. The so-called ``extreme He-B-stars'' and the ``He-sdO''
(post-AGB) stars are also He-dominated and could represent objects in
transition phases between RCB and O(He); see Tab.~\ref{tababu} for
parameters of some representatives.

The evolutionary link between these He-dominated post-AGB objects still needs to
be investigated. They are possibly the result of a merging process of two white
dwarfs \citep[e.g. ][]{saio:02} and, hence, these objects are not of
immediate interest to this review.

\subsection{Historical (very) late He-shell flashers}
\label{kap:hist}
Three stars have been identified as
actual born-again stars, mainly through their historical variability. These are FG Sge \citep{gonzalez:98}, V605 Aql
\citep{clayton:97} and V4334 Sgr \citep[Sakurai's object, ][]{duerbeck:96}.

For each of these cases an important question is whether the born-again
evolution is following a very late thermal pulse (VLTP) or just a late thermal
pulse (LTP). As we will discuss in \kaps{sec:vltp} and \ref{sec:ltp} the difference
between the two is that the VLTP happens late during the post-AGB evolution, and
induces a H-ingestion flash that consumes H in the surface layer by nuclear
burning. As recent evolution calculations show (\kap{sec:vltp}), the VLTP born-again star follows
a characteristic double loop in the HRD, where the first return to the AGB proceeds fast in
only a few years, whereas the second return takes much longer, of the order
$10^2\jahre$. The LTP happens earlier, on the horizontal,
constant luminosity part of the post-AGB track, and H-deficiency is only the
result of dredge-up mixing when the star returns to the AGB. The LTP born-again
star follows a single loop on a long time scale, similar to the second return of
a VLTP born-again star.

V605~Aql has experienced a VLTP in 1917
\citep{lechner:04}. As summarized in \citet[ see more references
there]{clayton:97} it brightened over a period of only two years, followed by
three episodes of fading and brightening. Then V605~Aql disappeared, enshrouded
in its own dust. In the early 1970s its position was found to coincide
with that of the planetary nebula Abell~58. It was later discovered that the PN in
fact has a small high-velocity, H-deficient central knot, while the outer nebula
is H-normal. Spectra taken in 1986 by \citet{seitter:87a,seitter:87b} reveal a broad stellar
\ion{C}{4} emission line indicative for a very hot (\Teff$\approx$100\,000~K)
[WC] central star, implying that V605 Aql
had already started to reheat and became a hot H-deficient or H-free central
star. A recent model atmosphere analysis of a VLT spectrum taken in 2002
confirms that V605~Aql is now a [WCE] star, having \Teff=95\,000~K \citep{fedrow:05}.

For FG~Sge it is not so obvious whether it is a LTP or a VLTP born-again
star. \citet{lawlor:03} proposed that while Sakurai's object and V605 Aql are
fast evolving VLTP first-return born-again objects, FG Sge has experienced a
VLTP too, but it is now on the second, long time-scale return. However, as
summarized by \citet[][ see more references there]{gonzalez:98}, FG Sge started
to brighten in 1894 until the mid-1970s. Since then the brightness has not
changed. FG Sge had initially (in the 1960s) solar abundance. Later (in the
1970s), rare earth elements appeared on the surface, and in 1992 the star
underwent dramatic photometric variations due to dust condensation. In the 1990s
the carbon and heavy element abundance has further increased, and there is now
evidence for H-deficiency. The fact that FG Sge had solar abundances 40 years
ago, and only rather recently has shown evidence for H-deficiency is a clear
indication that FG Sge is a LTP star as pointed out by \citet{gonzalez:98}. A
VLTP born-again star during the second loop should have been already extremely
H-deficient, or likely even H-free during the entire past observed evolution,
which is not the case. We note, however, that it has been questioned if FG~Sge
has really become H-deficient \citep{schoenberner:02}.

Sakurai's object is the most recently discovered born-again star
\citep{duerbeck:97,duerbeck:00}. Its VLTP has been dated semiempirically
between 1992 and 1994, and its \Teff\ dropped below $10\,000\kelv$ during early 1996. At
this time the star was already H-poor, and continued to change the surface
abundance distribution over a period of six months, covered by observations of
\citet{asplund:99a}. While the H-abundance was decreasing even further, the
already high Li abundance continued to increase. In addition, these observations
showed the C, N and O abundances to be enhanced significantly compared to solar, as
well as a low $\czw/\cdr$ and a low Fe/Ni ratio. Since these observations
Sakurai's object has shared the fate of the other born-again objects and faded
away behind thick clouds of dust. Recent radio observations show that Sakurai's
object has already started to reheat \citep{hajduk:05}, and it too will
eventually become a hot H-free or H-deficient central star.

The observations of born-again stars have revealed important
information about the mass loss of these objects. \citet{hajduk:05} find a mass
loss rate between $10^{-5}$ and maybe up to $2\cdot10^{-4}\msun/\jahre$ for
Sakurai's object during the coolest evolution phase. For FG Sge \citet{gehrz:05}
determined a mass-loss rate of $2.3 \cdot 10^{-5}$ to $1.2 \cdot 10^{-4}
\msun/\jahre$.

\section{Stellar evolution origin}\label{theory}

The abundance pattern observed in H-deficient bare stellar cores can be
understood in terms of the evolutionary origin of these stars. The H- and
He-burning shells of AGB stars are the nucleosynthetic origin of the material that is
observed at the surface of PG1159 and [WC]-CSPN. In order to make this
connection plausible we review in this section first those aspects of AGB
evolution and nucleosynthesis that are relevant for our purpose, followed by a
summary of the current understanding of the born-again evolution that leads to
almost complete H-depletion. In \kap{sec:modabund} we describe the surface
abundance predictions of H-deficient post-AGB stars.

\begin{figure*}
\begin{center}
\includegraphics[scale=.60]{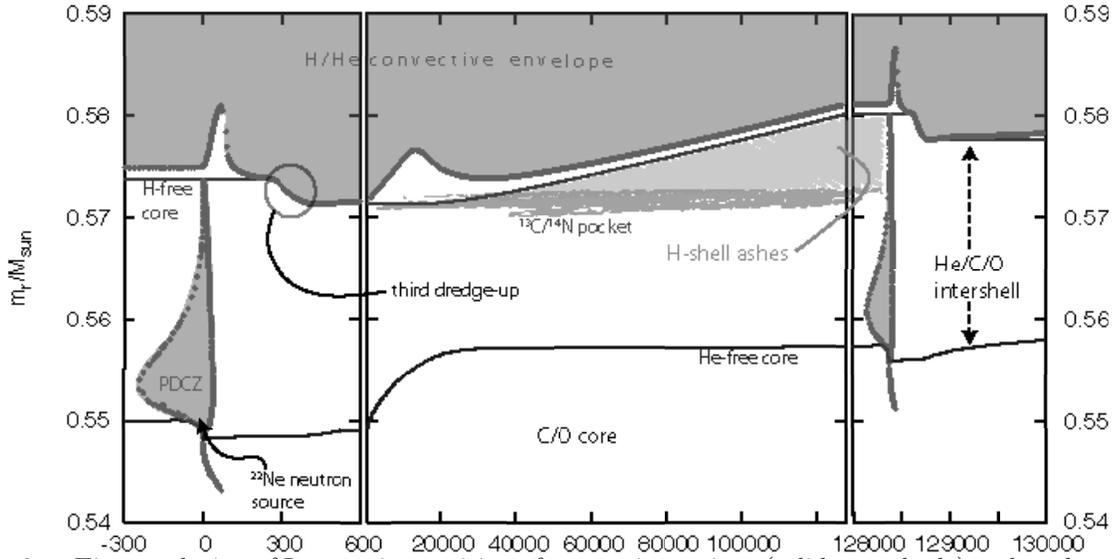}
\end{center}
\caption{\label{fig:kipp} Time evolution of Lagrangian position of
convective regions (solid grey shade) and nuclear burning shells (at the
location of the H- and He-free cores, respectively). The shown time interval comprises one
pulse-cycle. Each dot along the boundaries of convective regions corresponds to one time
step in the model sequence. }
\end{figure*}

\begin{figure}
\plotone{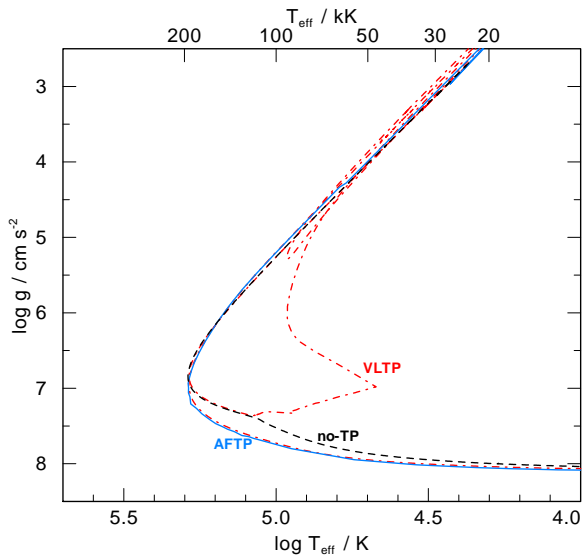}
\caption{Comparison of three different evolution channels in the
$g$--\Teff--plane for a $0.604\msun$ post-AGB star that had a main sequence mass
of $2\msun$ and evolved through about a dozen thermal pulses with
dredge-up. }\label{fig:ba-channels}
\end{figure}

\subsection{AGB evolution}
\label{sec:agbevolution}
The basic properties of AGB stars are covered in \citet{iben:83b}, AGB
properties specifically in view of the post-AGB evolution leading to H-deficient
cores have been summarized by \citet{bloecker:00a}, while an extensive review of
more recent developments has been provided by \citet{herwig:04c}. AGB stars have
an electron-degenerate C/O core as a result of core He-burning. Their nuclear
sources are the He- and the H-burning shells, that surround the core. Double-shell burning
on degenerate cores is unstable, and fundamentally similar to the X-ray bursts
on the surface of neutron stars. AGB stars experience quasi-periodic bursts of
the He-shell. These He-shell flashes or thermal pulses generate a peak
luminosity of several $10^8\lsun$. This large energy generation causes
convective instability of the layer between the He- and the H-burning
shells. This layer between the shells is the intershell. The convective
instability driven by the He-shell flash is the pulse-driven convection zone
(PDCZ). The recurrent thermal pulses, as well as the consequences for
nucleosynthesis and mixing are illustrated in \abb{fig:kipp}. Shown is the time
evolution of the narrow region (in Lagrangian coordinates) from the top of the
inert C/O core to the bottom of the envelope convection that reaches all the way
to the surface. The nuclear production in thermally pulsing AGB stars is determined
by the interplay of the PDCZ and the envelope convection. This interplay allows
the consecutive and alternating exposure of material to He- and H-burning. It
can be explained best by following the events of one full pulse cycle as shown
in \abb{fig:kipp} in detail.

\subsubsection{The thermal pulse cycle}
The He-burning that provides the energy for the PDCZ is dominantly the
triple-$\alpha$ reaction, producing \czw\ from \hevi. In addition a larger
number of additional $\alpha$-capture reactions are activated. Among these is
the $\nezw(\alpha,\n)\mgfu$ reaction that releases neutrons and leads to a large
neutron flux at the bottom of the PDCZ. We will get to the origin of \nezw\ in
the PDCZ in a moment. The neutron irradiation has two effects. It leads to the
production of many n-rich species heavier than iron, the \spr\ elements, and at
the same time it reduces the abundance of the seed species of the \spr, the
most important of which is \fese.

The H-shell at the bottom of the convective envelope reacts to the perturbation
of the He-shell flash in a characteristic way. Due to the expansion and cooling
of the layers above the He-shell, the H-shell is temporarily extinguished. After
briefly receding ($0 < t < 100\jahre$ in \abb{fig:kipp}) the convective envelope
engulfs more material from deeper layers. Eventually a significant amount of
material that has just been mixed and nuclear processed in the PDCZ is
convectively mixed into the envelope and to the stellar surface by this third
dredge-up\footnote{The first and second dredge-up can occur earlier during the
star's evolution after the H- and He-core burning ceases.}. At the end of the
third dredge-up the H-rich envelope and the \czw-rich intershell layer are in
direct contact, and any partial mixing, such as due to overshooting
\citep{herwig:97} or internal gravity waves \citep{denissenkov:02} leads to a
thin layer that contains both H and \czw. As this layer resumes contraction
after the thermal pulse and gradually heats up, H-shell burning eventually
starts again. As part of that process any partial mixing layer of H and \czw\
will produce an excess of \cdr. Detailed models have revealed that the neutron
release in the \cdr-pocket through the reaction $\cdr(\alpha,\n)\ose$ plays the
most important role for the \spr\ in AGB stars \citep{gallino:97b,busso:99}.

As the H-shell burns outward during the interpulse phase it leaves behind
H-burning ashes, which mainly consist of \hevi. However, any CNO material in the
envelope, including previously dredged-up \czw\ will be transformed into
\nvi. In addition the He-shell ashes contain fresh \fese\ from the
envelope. Just below the region with H-shell burning ashes is a thin layer that
contains the \cdr-pocket. The fact that the \cdr\ pocket contains a highly
\nvi-rich region just above the \cdr-abundance peak is extremely important in
models that include shear mixing induced by differential rotation
\citep{herwig:02a,siess:04}. The abundance distribution in the following PDCZ is
therefore a mix of the H-shell ashes, the s-process enriched material from the
nuclear processed \cdr-pocket, and material that has been mixed and exposed to
nucleosynthesis in the previous PDCZ. Obviously the previous PDCZ is similarly a
mix of these different components. Full stellar evolution calculations including
sufficiently detailed nuclear networks can generate quantitative predictions
for the intershell abundance distribution after many thermal pulses.

\subsubsection{Convective extra mixing and rotation}
\label{sec:cem}
Convective extra mixing refers to a complex family of physical processes that
lead to convection induced mixing in stable layers adjacent to a convectively
unstable zone. Although convective extra mixing is sometimes referred to as
simply overshooting, this term is reserved for the convective fluid motions that
extend from the unstable into the stable layers, and leave the thermodynamic
stratification sub-adiabatic. If convective plumes that reach into the stable
layers lead to an adiabatic stratification the convective extra mixing is
referred to as penetration \citep{zahn:91}. However, yet another physical process
can convectively induce mixing beyond the convective boundary. Convective
motions may perturb the convective boundary which can lead to the excitation of
internal gravity waves \citep{press:81}. Such gravity modes can induce mixing in
the radiative vicinity of convective boundaries
\citep{garcia-lopez:91,montalban:94,denissenkov:02}.

Over the past decade hydrodynamic convection simulations have generated
additional insight into the relative importance of these convective extra mixing
processes.  A very good example of overshooting has been observed in the shallow
surface convection simulations by \citet{freytag:96}. Convective motions, both
downdrafts and upflows cross the convective boundaries without effort. The
exponential decay of the convective motions stretches over a pressure scale
height. Due to the very soft convective boundary no gravity wave excitation
could be observed in these simulations. Hydrodynamic simulations by
\citet{bazan:98} and \citet{young:05b} have shown that the oxygen-shell convection in
massive stars is a different regime. Although overshooting is observed in these
calculations as well, an equally important or maybe even dominating process for
mixing is the excitation of gravity waves that generate turbulence in the
radiative layers next to the convective boundary. The excitation of gravity
waves is possible because the convective boundaries are much stiffer than in
shallow surface convection. Qualitatively similar results have been found for
He-shell flash convection in AGB stars by \citet{herwig:06a}. Although
convective extra mixing has not been analysed quantitatively, this study shows
that overshooting motions are severely limited in their ability to cross the
stiff convective boundary, especially the bottom boundary. However, the boundary
stiffness leads to the excitation of high-amplitude gravity waves. The most
recent simulations indicate that a small amount of mixing across the convective
boundaries of the PDCZ exists, both at the top and at the bottom
boundaries.

Convective extra mixing at the bottom of the PDCZ has been included in AGB
calculations as an exponential decay of the mixing-length theory convective
diffusion coefficient \citep{herwig:97,althaus:05a}. Such extra
mixing leads to higher temperature at the bottom of the PDCZ \citep{herwig:99a}
with implications for the \spr\ branching nucleosynthesis driven by the \nezw\
neutron source \citep{lugaro:02a,herwig:06a}. This limits the maximum amount of
convective extra mixing at the bottom of the PDCZ, but a detailed analysis to
quantify this limit is not yet available.  For the H-deficient post-AGB stars
another effect of this mixing is more important. The models predict that larger
convective extra mixing leads to a larger intershell abundance of \ose. As we
will explain below, the observed oxygen abundance of H-deficient post-AGB stars
are in very good agreement with AGB models that include this extra mixing
at the bottom of the PDCZ. The hydrodynamic properties of convective boundaries
are intimately related to the predicted abundance patterns of H-deficient
post-AGB stars.

The effect of stellar rotation, on the other side, has not yet been studied in
the same detail. Stellar models of AGB stars
including the effect of rotation have shown that the \spr\ nucleosynthesis
efficiency in the \cdr\ pocket may be reduced compared to non-rotating models
\citep{herwig:02a,siess:04}. Indirectly, convective extra mixing at the bottom
of the PDCZ could compensate for the reduced \spr\ efficiency in rotating AGB
models through a larger \czw\ intershell abundance \citep{herwig:06a}. While, as
we will discuss below, such a larger \czw\ intershell abundance is in agreement
with observations of H-deficient post-AGB stars, detailed models of this
scenario are not yet available. It therefore seems that stellar rotation at this
point is rather indirectly relevant for the interpretation of abundance patterns
of H-deficient post-AGB stars.

\subsection{Post-AGB evolution}

When the AGB star has lost all but approximately $10^{-2}\msun$ of its envelope
mass it starts the transition from the giant configuration to the white dwarf
configuration \citep{schoenberner:79}. The remaining envelope contracts and the
star evolves at constant luminosity to the hot stellar temperatures
(\Teff $>$ 30\,000~K) that are required for central stars to ionize the planetary
nebulae. Mass loss plays an important role in determining the transition
velocity during this phase \citep{bloecker:00a}. Eventually hydrogen
shell-burning stops and both the stellar luminosity and effective temperature start to
decrease (\abb{fig:hrd}). The star is about to enter the white dwarf cooling
track. Up to this point the star retains the ability to ignite a He-shell flash,
if the amount of He accreted from the H-shell is large enough, and the density
in the He-shell is high enough. Such a post-AGB He-shell flash will initiate a
born-again evolution during which the star is reborn as a giant star. The basic
concept of the born-again evolution scenario was established by \citet{iben:83a}
and \citet{schoenberner:83}. It is noteworthy that \citet{iben:83a} have already
speculated about the existence of two distinct born-again evolution channels, to
which we now refer as the late thermal pulse (LTP) and the very late thermal
pulse (VLTP). We add to those a variation of single star evolution that may be
able to produce (possibly only mildly) H-deficient hot post-AGB stars without a
born-again evolution, the so-called AGB final thermal pulse (AFTP).

\subsubsection{VLTP}
\label{sec:vltp}
In the VLTP born-again evolution case the PDCZ (\abb{fig:kipp}) is able to
penetrate into the remaining H-rich envelope, triggering a H-ingestion flash
(HIF) that imposes its own nuclear burning and mixing signature on the
subsequent evolution \citep{bloecker:00a,herwig:00a}.  This HIF is possible when
the He-shell flash occurs very late in the post-AGB evolution, when the star is
converging onto the white dwarf cooling track. At that time the H-shell is
inactive, and provides no entropy jump that could prohibit convective
instability below to spread outwards.

The time at which a He-shell flash can occur during the post-AGB evolution
depends on the thermal pulse phase at which the star leaves the AGB. 
A departure too early in the cycle means that no
He-shell flash will happen, a departure too late in the cycle means a He-shell
flash is triggered early in the post-AGB phase when the H-shell is still on,
resulting in a LTP born-again evolution, rather than a VLTP.

Most born-again models that have been presented more recently are in fact VLTP
models \citep{iben:95b}. The chain of events starts with the onset of the
He-shell flash convection that spreads over the entire intershell as the
He-burning luminosity quickly increases in thermonuclear runaway.  Close to the
peak luminosity the convective instability extends into the H-rich,
unprocessed envelope.  Protons are transported on the convective time-scale into
the hot, deep layers of the PDCZ. Eventually the proton-rich material reaches
temperatures that are high enough to allow the $\czw(\p,\gamma)\ndr$ reaction to
proceed on the convective time scale. At this position within the PDCZ the peak
H-burning energy will be released. If we believe the 1D-stellar evolution models
the energy generation leads to the formation of a separate convection zone
driven by the H-burning energy, and separated from the He-shell convection zone
beneath by a small radiative layer. The upper H-burning layer probably engulfs
the entire material up to the stellar surface, and burns its hydrogen
content. Whatever traces of H maybe left will very likely be shed off by mass
loss during the subsequent giant phase.

The VLTP model computations certainly are among the numerically more difficult
stages of stellar evolution. Nevertheless, several improvements have been made
in recent years. The VLTP model by \citet{herwig:99c} included a reliable, fully
implicit numerical method for solving simultaneously the nuclear network and
the time-dependent mixing equations. They also based their post-AGB evolution
sequence on an AGB progenitor with convective extra mixing (\kap{sec:cem}). In
that way the model could reproduce the observed high O abundance in
[WC] and PG1159 stars. For the surface abundances, especially oxygen,
\citet{althaus:05a} come to the same conclusion.

When it became clear that Sakurai's object was in fact a VLTP case in which the
born-again evolution from the pre-WD to the giant phase was witnessed in
real-time (\kap{kap:hist}) it also became clear that this star was not following
the script that had been laid out by any of the previous calculations. The
born-again evolution from the He-shell flash to the giant configuration has been
observed to be only about two years. However, any previous VLTP calculation
predicted born-again evolution times of a factor $10$ \citep{iben:95} to $100$
\citep{herwig:99c} longer. An important difference between these
two calculations is the numerical treatment of simultaneous burning and mixing,
opacities, and metallicity, which collectively make a factor of 10 difference
very plausible. For example, \citet[][ Table 2, section D-E]{lawlor:03} show that
the duration of the constant luminosity part of the born-again evolution is
about three times shorter for $Z=0.001$ compared to $Z=0.02$. In any case, the
important point is that these calculations do agree in the fact that Sakurai's object was
retracing the evolution back to the AGB much faster than predicted by VLTP
models. This indicated, that the born-again evolution time is sensitive to the
numerics, and as well to the physics of burning and mixing, and possibly other
input physics.

Subsequently a new generation of models of the VLTP has appeared. At the 2000
workshop \emph{Sakurai's object: What have we learned in the first five years?}
in Keele, UK, \citet{lawlor:02} presented VLTP tracks with two loops, similar to
the track shown dashed in \abb{fig:hrd}. However, the difference between these
new models with double-loop \citep[see also][]{2003IAUS..209..111H,hajduk:05}
and previous models was not clear initially. Then, \citet{herwig:01a} showed a
connection between the convective mixing speed of surface H into the hot He- and
\czw-rich layers deeper inside and the born-again evolution time from the hot
pre-white dwarf stage back to the giant configuration. 
The \citet{lawlor:02} and the \citet{herwig:01a} models as well as  all
later VLTP models use some simultaneous treatment of nuclear burning and
convective time-dependent mixing in which the nuclear network equation sets in
each mass shell are solved together with the diffusion equations for each
species. Although the basic concept is similar in all more recent models the
details of the numerical solution technique differ, with possible consequences
for the calculated born-again time scale.

\citet{herwig:01a} performed a series of tests for a $0.604\msun$ post-AGB star
with a range of mixing speeds. Peak nuclear burning of ingested H with \czw\
should occur where the nuclear and the mixing time scales are the same. As the
temperature strongly increases as protons convectively diffuse inward the
nuclear time scale decreases. For reduced mixing velocity the mixing time scale
is larger and the position of equality of nuclear and mixing time scales is
closer to the surface than with a faster mixing velocity as predicted by the
mixing-length theory. These models suggested that a solar metallicity model with
a mass of $0.604\msun$ could reproduce the born-again evolution time of
Sakurai's object if the convective mixing speed is reduced by a factor $100$.

\citet{lawlor:03} discuss the effect of reducing the mixing efficiency for three
values: reduction by $10^2$, $10^3$ and $10^4$. This may be a different regime
than the one explored by \citet{herwig:01a}, because they report a longer
evolution time scale for a reduction factor $10^4$ than for $10^3$. It should be
noted however, that neither of these models included some possibly important
physics, like the effect of $\mu$-gradients on convective boundaries, or
time-dependent convective energy transport.

The emerging picture, that was first fully described by \citet{lawlor:03}, is
that the VLTP evolution is a superposition of two flashes, reflected in the
double loop morphology of the HRD track. The first fast born-again evolution is
the result of the H-ingestion flash, whereas the second longer lasting loop is
driven by the He-shell flash that is still proceeding at the bottom of the
intershell, largely unperturbed.

Even more recently, \citet{althaus:05a} have presented new calculations of the
VLTP, that connect the VLTP event with the observational properties of PG1159
stars, and DB and DQ white dwarfs. These calculations confirm the double-loop
picture, and feature a born-again evolution time of $50\jahre$.
\citet{miller-bertolami:05} investigate the dependence of the born-again times
on additional physics, like the convection theory and the $\mu$-gradients, and
on numerical aspects, like time resolution. The born-again time of their best model
is $17\jahre$ without changing the mixing velocity \emph{ad hoc}. They do confirm the
relation between mixing velocity and born-again time of \citet{herwig:01a}.

\subsubsection{Discussion of the H-ingestion mixing during the VLTP}
As the discussion in the previous section shows, the more recent stellar
evolution studies of the VLTP show some agreement (e.g.\ the double-loop
structure) as well as disagreement (e.g.\ the quantitative born-again time
scale). All these calculations are based on some assumptions that are well
founded in many phases of stellar evolution: spherical symmetry, convection can
be described by time-averaged quantities as in mixing-length theory, hydrostatic
equilibrium. However, several or maybe all of these assumptions are invalid for
the VLTP. The ingestion of hydrogen into the He-shell flash convection zone is a
complicated multi-dimensional problem of turbulent convective-reactive fluid
flow in which composition changes and reactive energy release are coupled to the
dynamics \citep{dimotakis:05}.

There is considerable uncertainty in the treatment of non-reactive mixing at the H-rich -- C-rich interface at the
top of the convection zone. Convective rising plumes will perturb this
interface, possibly penetrating a small distance. The perturbation of the
boundary layer induces gravity waves \citep{herwig:06a} which cause, possibly
in the interaction with a small amount of convective penetration, shear flows
and thereby turbulent mixing at this interface. This boundary mixing will be
very important because the much lower molecular weight of material in the H-rich
stable layer will prevent unmixed blobs to be entrained easily into the He-shell
flash convection zone. Nevertheless, as any stellar convection simulation shows,
including those of the He-shell flash convection of \citet{herwig:06a}, the
overall convective flow morphology is dominated by structures that have roughly
the size of the vertical extent of the convection zone. On these large vertical
scales confined downflows will eventually contain material with a significant
H-abundance. Nuclear energy from the $\czw(\p,\gamma)\ndr$ reaction is released
in localized bursts at the vertical position where the temperature is large
enough to reduce the nuclear burning time scale down to the local mixing time
scale. This mixing time scale corresponds to the actual velocity in the
downflow, not the rms-averaged value, that corresponds in fact well with the
mixing-length velocity.

In most stellar evolution situations the nuclear time scale is much longer than
the convective time scale. This justifies the customary operator split between
the convective mixing step and the nuclear burning calculation. It also justifies
the assumption that nuclear energy release is isotropic in the horizontal
direction (or on spheres). However, in the H-ingestion flash there are small and
irregular parcels of energy production that coincide with the horizontal area of
H-rich downflows. This breaks spherical symmetry.

The bursts of energy will locally add buoyancy and alter the convective flow
patterns, deflecting focused downflows sideways or even reversing the
flows. This has its correspondence to the notion derived from 1D models that the
energy release from the H-ingestion flash will lead to a split of the convective
region associated with an impenetrable entropy barrier. Whether or not such a
separation of convection zones will actually develop in real stars, and whether
the 1D mixing-length theory-based diffusion-burning models can account for this
evolution phase at all is uncertain, considering the many unknown aspects of the
complicated reactive, turbulent mixing that dominates the above sketched picture
of the problem. This picture is based on our preliminary multi-dimensional
hydrodynamic calculations of the H-ingestion problem.

In view of all these complications it seems rather surprising that the more
recent 1D VLTP stellar evolution models agree to some extent with each other,
and with the observed evolution of Sakurai's object at all. However, there is
room for some doubt whether the current models describe the physics correctly on
the level needed to understand the observations.

\subsubsection{LTP and AFTP}\label{sec:ltp}
If the He-shell flash ignites during the earlier horizontal, constant-luminosity
evolution the star will follow a born-again evolution but without a hydrogen
ingestion flash (HIF). This
variant of the post-AGB He-shell flash is the late thermal pulse
\citep{bloecker:00a,bloecker:97}. Although the LTP does not deplete H in the
surface layer by nuclear burning it will still develop a H-deficient surface
composition \citep{herwig:00a}. This is achieved entirely by mixing. As the star
inflates and the surface becomes cooler, envelope convection, that is
characteristic for giant stars, emerges again. At this point a dredge-up event in
the same way as following the AGB thermal pulses will mix a very thin H-rich
envelope layer of the order $10^{-4}\msun$ with a few $10^{-3}\msun$ of H-free
intershell layer. The result is H-deficiency as a result of dilution. The LTP
model of \citet{herwig:00a} predicts a H surface mass fraction of $X_\mem{H} =
0.02$.

The LTP has only a single loop back to the AGB which proceeds on a long time
scale of the order $10^2\jahre$. In addition there are differences in the
abundance evolution resulting from the fact that the LTP did not experience a
H-ingestion flash (\kap{sec:modabund}). These include no Li production, no
enhanced \nvi\ abundance, a larger $\czw/\cdr$ ratio than in the VLTP HIF
models. Unfortunately there are no detailed LTP models including all relevant
nucleosynthesis and a realistic treatment of the AGB progenitors available.

The born-again scenario may imply that
H-deficient CSPN should have systematically older PN, which is not the case
\citep{gorny:01}. In contrast, the appeal of the AFTP scenario is that it does not
require the central star to evolve through a lengthy first CSPN phase and the
return to the AGB and to the CSPN again.  The AFTP leads to H-deficiency at the last
thermal pulse still on the AGB. By invoking some fine-tuning of the mass loss
the envelope mass at the last AGB thermal pulse is so small that the star leaves
the AGB immediately after the dredge-up following this last AGB thermal
pulse. Because of this very small envelope mass the dredge-up leads as in the
LTP to a significant dilution of the H surface abundance.

The AFTP scenario is in contradiction to AGB stellar evolution studies of
dredge-up towards the end of the AGB evolution, which assert that dredge-up
stops when the envelope mass falls below a certain value. For example, a minimum
envelope mass of $0.5\msun$ for dredge-up was reported by
\citet{straniero:97}. However, dredge-up predictions in stellar evolution models
are notoriously dependent on fine-tuning some mixing parameters, like the
mixing-length parameter or the overshooting parameter.

The two AFTP models of \citet{herwig:00a} were calculated with a mixing-length
parameter of $\alpha_\mem{MLT}=3.0$ and overshooting. At the time of the last
AGB thermal pulse the envelope mass was $3\cdot10^{-2}\msun$ and
$4\cdot10^{-3}\msun$, and the dredged-up mass was in both cases
$6\cdot10^{-3}\msun$. This resulted in H-abundances at the surface of the
post-AGB star of $X_\mem{H}=0.55$ and $0.17$, respectively. This is much larger
than the H-abundance observed in PG1159 stars. However, the AFTP model may
explain quite naturally the ``hybrid'' PG1159 stars which are H-deficient but yet
have H abundances much larger than the typical PG1159 and [WR]-CSPN stars
\citep[Tables~\ref{tababu} and~\ref{tabpg1159};][]{napiwotzki:91}.

The main problem with the AFTP scenario seems to be a missing mechanism that
favors the pulse phase zero (or very small) for the departure from the
AGB. Somehow, the thermal pulse must trigger enhanced mass loss that in turn
enhances the probability of a departure at phase close to zero. The typical AGB
mass-loss formulae do not provide such a mechanism and would lead to a very low
probability of such an event. 

\citet{demarco:02b} considered the possibility that swallowing a planet or a
very-low mass star companion could eject the star with high probability during
the temporary, significant radius increase induced by the thermal pulse.
The multi-dimensional simulations showed that envelope ejection is indeed
possible. This would initialize the departure from the AGB. 
Clearly this interesting possibility needs more investigation, and such
efforts are underway.

\subsubsection{The HRD tracks of post-AGB evolution and the importance of the AGB progenitor evolution}
\label{sec:hrdnew}

The evolution tracks of an undisturbed H-normal post-AGB evolution (no-TP), an
AFTP track and a VLTP track are compared in \abb{fig:ba-channels}. All three
tracks have been started from the same tip-AGB model that was evolved through
all previous evolution phases from a $2\msun$ main-sequence model. The stellar
mass at the time of departure from the AGB is $0.604\msun$. The comparison shows
that surface abundance has no effect before the ``knee'' (the hottest point in
the track) and only a small effect after the knee, where the H-normal track is
slightly cooler for a given $\log g$. However, this difference should not be
overemphasized, because it is largely due to a ``kink'' in the track that
reflects the onset of the reignition of the He-shell. In the no-TP track the
He-shell flash run-away does not develop. For the HRD track after the knee it
seems not to matter how the star became H-deficient. The AFTP and the final CSPN
evolution of the VLTP have identical tracks. This indicates that H-normal CSPN
tracks can be used as a tool to determine masses from spectroscopic stellar
parameters in the absence of a homogeneous grid of H-deficient tracks.

Another point that is evident from the comparison of tracks in \abb{fighrd}
seems to be more important for accurate mass determinations. This figure shows
older tracks from \citet{schoenberner:83}, \citet{bloecker:95b} and
\citet{wood:86}, 
as well as our newer VLTP track that is also shown in
\abb{fig:ba-channels}. While the $0.605\msun$ track from \citet{bloecker:95b} and
the $0.6\msun$ track from \citet{wood:86} show good agreement the new
$0.604\msun$ track is at and below the knee $\Delta \log \teff = 0.1 - 0.15$
hotter. This is even more surprising since the same code (although with some
differences concerning nucleosynthesis and mixing) has been used for the $0.605$
and the $0.604\msun$ model.

The reason for this HRD track difference is a different dredge-up or mass-loss
history or both in the AGB progenitor evolution. For the new $0.604\msun$ track
the AGB progenitor evolution includes a complete sequence of thermal pulses, a
large number of which are followed by third dredge-up mixing events. These
dredge-up events lead to a smaller effective core growth, while the properties
of the core itself, in particular the radius, depends to a large extent on the
increasing degeneracy of the C/O core \citep{herwig:98b,mowlavi:99}. If an AGB
evolution sequence experiences many efficient third dredge-up events the
resulting post-AGB star appears in the HRD like a slightly more massive star
compared to the same core mass post-AGB stellar model that has an AGB thermal
pulse history of inefficient or no dredge-up.

Mass loss can have a similar effect because it determines the age of the core
for a given initial mass at the tip of the AGB \citep{bloecker:90}. This was
shown in detail by \citet[][ his Fig.\,10]{bloecker:95b} who calculated two post-AGB
tracks with identical mass ($0.84\msun$), one with a $5\msun$ main-sequence
progenitor, and one with initially $3\msun$. In order for the $3\msun$
model to reach such a high core mass, a smaller mass-loss rate than for the
$5\msun$ track was adopted. In this way the core had a long time to grow through
shell burning before the envelope loss forced the AGB departure. The result is a
more degenerate, more compact, and consequently hotter post-AGB stellar model
with the $3\msun$ progenitor compared to the post-AGB model with the $5\msun$
progenitor.

In summary, if CSPN tracks are computed for the purpose of determining masses
(or fading times) the third dredge-up and the mass loss during the progenitor
evolution must be accurately taken into account.

\subsection{Surface abundance predictions for H-deficient post-AGB stars}
\label{sec:modabund}
If we interpret H-deficient post-AGB stars, like PG\,1159 stars or [WC]-CSPN as
the bare cores of former AGB stars exposed by the born-again evolution or an
AFTP evolution, then the surface abundance of these stars is a superposition of
the intershell abundance of the progenitor AGB star at the last thermal pulse on
the AGB and any modification that in particular the violent burning and mixing
associated with the H-ingestion flash of the VLTP may have caused. In the
following we discuss the surface abundance predictions for the elements that
have been observed so far, taking into account both of these contributions.

\subsubsection{Carbon and Oxygen}
During the He-shell flash the triple-$\alpha$ reaction that transforms
three \hevi\ into one \czw\ is the dominant source of energy. The
layers immediately below the PDCZ contain an amount of \ose\ from the
$\czw(\alpha,\gamma)\ose$ reaction, increasing with depth. Starting
with the earliest models of thermal pulse AGB stars
\citep{schoenberner:79} it has been consistently found that without
convective extra mixing the \ose\ abundance is 0.01--0.02, and
the \czw\ abundance is 0.2--0.25 in the intershell with the
remaining $\sim 0.75$ \hevi.

More recently models with convective extra mixing in the form of a parameterized
exponential overshoot at the bottom of the PDCZ have been constructed
\citep{herwig:97,herwig:99a,althaus:05a}. These models agree quantitatively that
such extra mixing leads to a systematically larger \ose\ and \czw\
abundance in the PDCZ.  \citet{herwig:99a} showed that the efficiency of
convective extra mixing is proportional to the intershell \ose\ abundance. In
addition, there is an evolution of the \ose\ abundance with pulse number,
initially rising, then reaching a maximum value after a few thermal pulses and
decreasing towards a plateau at the last thermal pulses. For the evolution
sequence with initially $3\msun$ \citet{herwig:99a} finds at the last computed
thermal pulse an intershell abundance of $X(\ose)=0.18$ and $X(\czw)=0.41$. For
a sequence with initially $2.7\msun$ \citet{althaus:05a} find $X(\ose)=0.23$ and
$X(\czw)=0.50$. Presumably this difference is due to the smaller number of
thermal pulses in \citet{althaus:05a} compared to \citet{herwig:99a}. As
mentioned above the \ose\  intershell abundance goes through a maximum
(corresponding to a minimum for \hevi) before approaching a plateau. The
\citet{althaus:05a} intershell abundance may correspond to a thermal pulse closer
to that maximum.

\subsubsection{Nitrogen}
Nitrogen is produced in AGB stars by H-shell burning on envelope
material enriched by carbon dredge-up after previous thermal
pulses. In low-mass AGB stars the H-burning shell burns most CNO
material in the envelope material it consumes, into \nvi. Therefore,
the H-shell burning ashes (see \abb{fig:kipp}) contain with each dredge-up pulse
more \nvi, that is mixed in the next He-shell flash
convection zone. In the He-burning shell during the flash
this \nvi\ is destroyed and forms \oac\ and
\nezw\ (see below). The models predict that the intershell of AGB
stars at the end of their evolution is practically free of \nvi.

In massive AGB stars \nvi\ is produced in the envelopes by hot-bottom
burning \citep{sackmann:92}. The dredged-up carbon in the envelope is
transformed by two p-captures into \nvi\ because part of the H-burning
shell is included in the convective envelope. Therefore, we expect in
the more massive post-AGB stars of roughly solar metallicity a
nitrogen mass fraction of $1$\% to maybe a few $\%$ in the H-rich
envelope at the time of departure from the AGB.

The nitrogen evolution is different for the LTP and VLTP. The LTP will
only dilute some intershell material with the remaining small amount
of envelope material. Even if the progenitor has evolved through the
hot-bottom burning phase, the nitrogen abundance will be at most
roughly $0.1\%$. If the star evolves through a VLTP, nitrogen is
produced by the H-ingestion and may be as high as a few $\%$. 
The absence or presence of N should be a reliable indicator of a
LTP or VLTP event.

\subsubsection{Neon}
During each interpulse phase the H-shell builds a layer of ashes, that
is engulfed into the following PDCZ. These H-shell ashes contain mainly \hevi,
with an important modification of the CNO abundances. All CNO, be it from the
stellar initial abundance or dredged-up after a previous thermal pulse, will be
mostly transformed into \nvi. This \nvi\ is exposed to the high He-burning
temperature ($T\approx 3\cdot 10^8\kelv$) where two $\alpha$ captures lead to
the formation of \oac\ first and then \nezw. Depending on mass some \nezw\  is
destroyed by the neutron producing reaction $\nezw(\alpha,\n)\mgfu$. For
low-mass AGB stars with initial masses around $1.5\msun$ the \nezw\ depletion is
only small ($\approx3\%$) because the peak temperature at the base of the PDCZ is
not reaching high enough values $>3\cdot10^9\kelv$.  Stellar evolution models
agree that the mass fraction of \nezw\ is about $2\%$. A certain spread in the
observed values is expected. The production of \nezw\  is larger in cases with
more dredge-up of \czw. The destruction of \nezw\ through $\alpha$-capture
increases with initial mass. As a trend the intershell \nezw\ should decrease
with increasing initial stellar mass.

\subsubsection{Iron and Nickel}
Fresh \fese\ is added to the intershell with the H-shell ashes, which are
processed envelope material and therefore contain the Fe-abundance of the
envelope. During the PDCZ \fese\ will be reduced by neutron capture and
transformed into Ni if the \nezw\ neutron source is active, i.e. \fese\
reduction will be more efficient in higher core-mass AGB stars. However, as it is
the case for many other species, the evolution of \fese\ in the intershell with
increasing pulse number depends sensitively on the dredge-up history, which in
turn seems in model calculations to be related to assumptions about convective
extra mixing \citep{mowlavi:99}.

Preliminary models by \citet{2003IAUS..209...85H} show that the \fese\ depletion
in the PDCZ at the last thermal pulse of a $3\msun$ star varies significantly
depending on stellar evolution model assumptions. For a thermal pulse AGB
sequence with convective extra mixing, showing very efficient third dredge-up
and high peak temperatures at the base of the PDCZ, the \fese\ depletion is
approximately $0.7\mem{dex}$, whereas a model without convective extra mixing
predicts a \fese\ depletion of only $0.2\mem{dex}$. In both cases all heavier
isotopes of Fe and Ni are significantly overabundant. An additional depletion of
Fe and production of Ni may be obtained during the late thermal pulse
(\kap{sec:vltp}). In any case, the unmistakable signature of Fe depletion due to
n-capture nucleosynthesis is the simultaneous increase of Ni, and
correspondingly a decrease of the Fe/Ni ratio. While the solar ratio
is about 20, material that has experienced a significant neutron
exposure would show a ratio closer to the \spr\ quasi steady-state of $\approx 3$.

\subsubsection{Fluorine}
\citet{jorissen:92} have observed high enhancements of fluorine in AGB
stars. They found a correlation of F enhancement and C/O ratio, which is a
strong indication that F is produced in the intershell and mixed to the surface
during the third dredge-up. Observations show that thermal pulses cause a
10-fold increase in \fne.

The nucleosynthesis path to \fne\ first involves the production of \nfu\ which
requires a neutron source:
$\nvi(\n,\p)\cvi(\alpha,\gamma)\oac(\p,\alpha)\nfu$. This chain of reaction
occurs during the interpulse phase. The neutrons are provided both by \cdr\ in
the H-shell ashes, and in the \cdr\ pocket. For a
certain range of mixing efficiencies, rotation induced mixing may enhance the production of \nfu\
\citep{herwig:02a}.  \nfu\ produced in either way is then engulfed by the
PDCZ. F is produced by $\nfu(\alpha,\gamma)\fne$.  The limiting factor in this
scenario is the efficient $\fne(\alpha,\p)\nezw$ reaction.  \fne-production
exceeds destruction only in a narrow temperature range at the base of the PDCZ,
between $2.2$ and $2.6 \cdot 10^8\kelv$ \citep{mowlavi:96}.

\citet{lugaro:04a} show that the \fne\ abundance in the He-intershell after the
last thermal pulse has a maximum between $2$ and $3.5\msun$ depending on
metallicity. For $Z=0.02$ and $Z=0.008$ the intershell abundance reaches a
maximum between $150$ and $290$ times solar ($X(\fne)_\odot = 4.1\cdot 10^{-7}$)
at $3.5$ and $3.0\msun$, respectively. However, for cores with larger or smaller
initial mass the \fne\ intershell abundance may be as small as the solar value.

Within an evolution sequence of thermal pulses for a given initial mass, and
even within an individual thermal pulse the \fne\  intershell shows a
considerable spread. The variation from pulse to pulse means that a quantitative
comparison with observed \fne\ values in H-deficient post-AGB stars depends on
details of the entire AGB evolution, including for example an accurate
description of mass loss and dredge-up to correctly model the total number of
thermal pulses. The spread within one He-shell flash convection event means that
there may be observable differences in the \fne\ abundance between VLTP with
H-ingestion and LTP born-again evolutions.

Finally it should be mentioned that \citet{lugaro:04a} find that nuclear reaction
rate uncertainties, in particular of the $\cvi(\alpha,\gamma)\oac$ and the
$\fne(\alpha,\p)\nezw$ reactions, significantly limit the accuracy of stellar
model predictions of \fne.

\subsubsection{Si, P and S}
These three elements have a too high Coulomb barrier to be altered by charged particle reactions
in AGB stars. This leaves only neutron captures to be considered for elemental
abundance shifts of Si, S and P. It is well known from the study of pre-solar
grains, that the neutron fluxes in AGB stars modify the ratios of the three
stable Si isotopes \citep{lugaro:99}. However, the elemental abundance of Si is
dominated by the lightest isotope, \siac, which in the solar abundance
distribution accounts for approximately $90\%$ of the elemental Si
abundance. Neutron exposure in AGB stars will shift a small fraction of \siac\
to \sine\ and \sidr, but the elemental abundance changes only very little in the
He intershell of AGB stars. Therefore, a solar Si abundance is expected in hot
H-deficient post-AGB stars.

For P the model calculation with initially $3\msun$ predicts an enhancement in
the intershell at the end of the AGB that is sensitive to the assumptions about
convective extra mixing. The expected range is from four times solar without
extra mixing to $\sim25$ times solar with extra mixing. As
discussed in \kap{sec:cem} the respective mixing algorithm contains a free
efficiency parameter. We consider the efficiency
chosen in this  model an upper boundary. As it is the case for \fne, the He-shell
flash nucleosynthesis can be sensitive to the initial mass. A systematic
evaluation of the P intershell abundance as a function of mass
and metallicity is not yet available.

The same models predict a small deficiency in S in the intershell at the end of
the AGB, ranging from 0.6 times solar with convective extra mixing to 0.9 times
solar for the standard model. A more systematic evaluation of the intershell
abundance evolution of S and P should include the neutron capture cross section
uncertainties. However, in general the uncertainties for these stable isotopes
are typically less than $10\%$ \citep{bao:00}.

\subsubsection{Lithium}
Although Li can not be observed in hot post-AGB stars, the observed Li
overabundance in Sakurai's object is an important hint. The production of this
fragile element in the form of its heavy isotope \lisi\ is usually
associated with hot-bottom burning in massive AGB stars
\citep{scalo:75,sackmann:92}. Enhanced extra mixing in red giant branch stars
induced by tidal synchronization or swallowing of a giant planet has been
proposed as another source of \lisi\ \citep{denissenkov:04a}.  Both of these
production sites are based on the \besi\ transport processes. \hedr\ and \hevi\
form \besi\ which becomes \lisi\ through a largely temperature
insensitive $\emi$-capture on a time-scale of approximately one year. If the
production region of \besi\ is convectively connected to cooler layers, like in
envelope convection that reaches into the H-burning shell in hot-bottom burning,
then \besi\ is transported out of the hot layers, and the resulting \lisi\ can
survive, because it is not immediately destroyed by $\p$-captures.

\citet{herwig:00f} have shown that if \hedr\ in the envelope has not been destroyed
by one of the two processes mentioned above (which is likely for most low-mass
AGB stars with initial masses below $\sim 3.5\msun$), then copious amounts of
\lisi\ can be produced in the hydrogen ingestion flash. They refer to this
process as hot H-deficient \hedr\ burning. The reason is somewhat different from
the \besi\ transport mechanism, in which \besi\ is transported out of the hot
p-rich, and therefore \lisi\ hostile environment quickly enough.  In the HIF the
amount of ingested H is strictly limited, and in addition to H, \hedr\ is
ingested into the He-shell flash convection zone. This \hedr\ quickly reacts
with \hevi\ and forms \besi, which then has to wait for an $\emi$ capture to
become \lisi. This waiting period is long enough for all the protons to be
rapidly consumed by \czw. By the time \lisi\ appears no protons are available
anymore for Li destruction, leading to a net production of \lisi. Current models
only account qualitatively for the processes involved. In a one-zone model
\citet{herwig:00f} find a maximum mass fraction of $X(\lisi)= 3\cdot10^{-6}$,
which corresponds to $\sim2.5 \mem{dex}$ more than the solar system meteoritic
value. We expect that the surface overabundance of Li in a realistic multi-zone
model would be much smaller.

\lisi\ production in a HIF has been recently confirmed by \citet{iwamoto:04} in
the context of He-shell flashes in extremely metal-poor AGB stars. In this
context it is interesting to note that \citet{cameron:71} in fact originally
proposed the \besi\ transport mechanism motivated by the first HIF models by
\citet{schwarzschild:67} that were, however, not reproduced by subsequent models
that included radiation pressure.

\section{Comparison of observation and theory}

In the previous Sections~\ref{observation} and \ref{theory} we have presented in
detail the element abundances observed in PG1159 and [WC] stars as well as the
the surface abundances predicted from stellar evolution models. How do they
compare?

Hydrogen: The observed H deficiency is in accordance with evolution models of
the born-again scenario. In the VLTP models H is ingested and burned and
disappears completely. An LTP event causes mixing of the H-rich envelope with
the intershell so that H is diluted down to a mass fraction of the order
0.02. This is close to the observational limit. The relatively high H abundances
in the hybrid-PG1159 stars and some [WC] stars (of the order 0.15) are explained
by AFTP models.

Helium, carbon and oxygen: These are the main constituents of the intershell
region of AGB stars and the model abundances generally agree with the observed
abundances in PG1159 and [WC] stars. The spread of observed relative
He/C/O ratios can in part be explained by differences in stellar mass and the
different number of thermal pulses experienced by objects even with equal total
mass (the latter being a consequence of potentially different AGB
mass-loss, for example as a result of rotation). However, we think that some few extreme cases of He/C/O ratios
remain unexplained, in particular objects with relatively low O abundance
(of order 0.01) and simultaneously high C abundance (of order 0.5). 
One speculative scenario has been sketched by 
\citet{demarco:03a}. If the AGB star has a low-mass stellar or planetary companion at a
distance of 1--2~AU it would be engulfed during the radius peak induced by the
first thermal pulse. The transfer of orbital energy into the AGB envelope may
eject the envelope and lead to a departure from the AGB. According to
\citet{herwig:99a} the intershell O abundance is very low after the first
thermal pulse even if overshooting has been applied. If such an evolution later
involves a VLTP or LTP, a PG1159 star with a very low O abundance would result. A
similar evolution could result from a low-mass star that naturally leaves the AGB just
before the first thermal pulse \citep{bertolami:06}.

Isotopic ratios can not be measured in hot post-AGB stars. However,
both the \czw/\cdr\ and the C/N ratios of our VLTP model are in good
agreement with the observations of Sakurai's object.

Nitrogen: The observations show a wide range of N abundance, from very
low upper limits ($<3 \cdot 10^{-5}$ in HS1517+7403) to mass fractions of $1 - 2\%$, in
agreement with predictions for LTP and VLTP evolution,
respectively. Intermediate abundance levels could in some cases be
associated with higher stellar mass. 

Neon: The observed high Ne abundances (of the order 0.02) agree well with the
model predictions.

Fluorine: The rich diversity of F abundances observed in PG1159 stars (from
solar up to 250 times solar) is explained by models with different stellar mass
leading to different F production efficiencies.

Silicon: The Si abundance in the evolution models is almost unchanged so that a
solar abundance is expected. Only few abundance analyses are available. PG1159
stars display the expected Si abundance while surprisingly strong overabundances
(up to 40 times solar) were found in some [WCL] stars.

Phosphorus: Theoretical predictions and observational data are
scarce. Current models are consistent with P overabundances of up to
4--25 times solar. Preliminary results from two PG1159 stars indicate
roughly solar abundances.

Sulfur: Preliminary abundance determinations in PG1159 stars suggest a
wide spread, ranging between 0.01--1 times solar. These results need
to be confirmed by further analyses.  Like in the case of P,
systematic investigations with evolution models are lacking, but
current models predict only slight (0.6 solar) S depletion.

Lithium: The Li abundance cannot be determined in hot objects so that
no conclusions can be drawn from PG1159 and [WC] stars. Sakurai's
object, however, shows a large Li abundance. As discussed in previous
sections this Li can not have survived from an earlier phase, and is a
strong additional hint that Sakurai's object did in fact evolve
through a VLTP.

Iron and Nickel: The observed Fe deficiency in PG1159 and [WC] stars is
explained by neutron captures on $^{56}$Fe. The consequent increase of the Ni
abundance cannot be confirmed or contradicted with current observational
material. However, Sakurai's object indeed exhibits a strongly subsolar Fe/Ni
ratio, that is quantitatively in agreement with model estimates.

To conclude, element abundances predicted by born-again evolutionary
models (LTP and VLTP case) as well as the AGB Final Thermal Pulse
models (AFTP case) can explain most observational results from PG1159
and [WC] star analyses.  The model surface abundances largely reflect
the intershell abundance of the AGB progenitor. From the
good qualitative and quantitative agreement between observed and predicted abundances we conclude that
[WC] and PG1159 stars generally display on their surface the intershell
material of the AGB progenitor. For this reason, detailed
quantitative spectroscopic investigations of these hot post-AGB stars can be
used as a tool to study the nuclear production site of the
He-intershell in AGB stars. Such a tool is in particular useful
because the AGB interior is otherwise obviously inaccessible to direct
observation. It is hoped that new observations and analyses of these H-deficient post-AGB stars
will contribute to an even more complete picture of the
nuclear astrophysics processes in AGB stars.

\acknowledgements F.H. would like to thank Lars Koesterke for his
early support in numerical methods for burning and mixing in 1D
born-again stellar evolution models. He would also like to thank
Bernd Freytag for their delightful collaboration on stellar interior
convection hydrodynamic simulations, and Orsola De~Marco for many
enlightning discussions about CSPN and other things. K.W. thanks
Stefan Dreizler, Uli Heber, and Thomas Rauch for their enduring close
collaboration. We would like to thank Maria Lugaro for her help with
preliminary estimates on P and S abundance predictions. Comments on an earlier
version of this paper by Orsola De~Marco, Wolf-Rainer Hamann, Thomas Rauch, and
Detlef Sch\"onbernber are gratefully acknowledged. This work was
funded in part under the auspices of the U.S.\ Dept.\ of Energy under
the ASC program and the LDRD program (20060357ER) at Los Alamos
National Laboratory. It was also funded in part by the German Science
Foundation (DFG) and the German Aerospace Center (DLR).  \abb{fig:hrd}
and \abb{fig:kipp} with minor modifications are reprinted, with
permission, from the Annual Review of Astronomy and Astrophysics,
Volume 43 (c) 2005 by Annual Reviews www.annualreviews.org.

%\bibliography{astro}

\begin{thebibliography}{146}
\expandafter\ifx\csname natexlab\endcsname\relax\def\natexlab#1{#1}\fi

\bibitem[{Acker \& Neiner(2003)}]{acker:03}
Acker, A. \& Neiner, C. 2003, A\&A, 403, 659

\bibitem[{{Althaus} {et~al.}(2005{\natexlab{a}}){Althaus}, {Miller Bertolami},
  {C{\'o}rsico}, {Garc{\'{\i}}a-Berro}, \& {Gil-Pons}}]{althaus:05b}
{Althaus}, L.~G., {Miller Bertolami}, M.~M., {C{\'o}rsico}, A.~H.,
  {Garc{\'{\i}}a-Berro}, E., \& {Gil-Pons}, P. 2005{\natexlab{a}}, A\&A, 440,
  L1

\bibitem[{{Althaus} {et~al.}(2005{\natexlab{b}}){Althaus}, {Serenelli},
  {Panei}, {C{\'o}rsico}, {Garc{\'{\i}}a-Berro}, \&
  {Sc{\'o}ccola}}]{althaus:05a}
{Althaus}, L.~G., {Serenelli}, A.~M., {Panei}, J.~A., {C{\'o}rsico}, A.~H.,
  {Garc{\'{\i}}a-Berro}, E., \& {Sc{\'o}ccola}, C.~G. 2005{\natexlab{b}}, A\&A,
  435, 631

\bibitem[{{Asplund} {et~al.}(2000){Asplund}, {Gustafsson}, {Lambert}, \&
  {Rao}}]{asplund:00}
{Asplund}, M., {Gustafsson}, B., {Lambert}, D.~L., \& {Rao}, N.~K. 2000, A\&A,
  353, 287

\bibitem[{{Asplund} {et~al.}(1999){Asplund}, {Lambert}, {Kipper}, {Pollacco},
  \& {Shetrone}}]{asplund:99a}
{Asplund}, M., {Lambert}, D.~L., {Kipper}, T., {Pollacco}, D., \& {Shetrone},
  M.~D. 1999, A\&A, 343, 507

\bibitem[{Bao {et~al.}(2000)Bao, Beer, K\"appeler, Voss, Wisshak, \&
  Rauscher}]{bao:00}
Bao, Z.~Y., Beer, H., K\"appeler, F., Voss, F., Wisshak, K., \& Rauscher, T.
  2000, ADNDT, 76, 70

\bibitem[{Barlow \& Hummer(1982)}]{barlow:82}
Barlow, M.~J. \& Hummer, D.~G. 1982, in Wolf-Rayet Stars: Observations,
  Physics, Evolution, ed. C.~de~Loore \& A.~Willis, IAU Symp. 99 (Dordrecht:
  Reidel), 387

\bibitem[{{Bauer} \& {Husfeld}(1995)}]{bauer:95}
{Bauer}, F. \& {Husfeld}, D. 1995, \aap, 300, 481

\bibitem[{{Bazan} \& {Arnett}(1998)}]{bazan:98}
{Bazan}, G. \& {Arnett}, D. 1998, ApJ, 496, 316

\bibitem[{Bl\"ocker(1995)}]{bloecker:95b}
Bl\"ocker, T. 1995, A\&A, 299, 755

\bibitem[{{Bl{\"o}cker}(2001)}]{bloecker:00a}
{Bl{\"o}cker}, T. 2001, \apss, 275, 1

\bibitem[{Bl\"ocker \& Sch\"onberner(1990)}]{bloecker:90}
Bl\"ocker, T. \& Sch\"onberner, D. 1990, A\&A, 240, L11

\bibitem[{Bl\"ocker \& Sch\"onberner(1997)}]{bloecker:97}
---. 1997, A\&A, 324, 991

\bibitem[{{Bond} {et~al.}(1996){Bond}, {Kawaler}, {Ciardullo}, {Stover},
  {Kuroda}, {Ishida}, {Ono}, {Tamura}, {Malasan}, {Yamasaki}, {Hashimoto},
  {Kambe}, {Takeuti}, {Kato}, {Kato}, {Chen}, {Leibowitz}, {Roth}, {Soffner},
  \& {Mitsch}}]{bond:96}
{Bond}, H.~E., {Kawaler}, S.~D., {Ciardullo}, R., {Stover}, R., {Kuroda}, T.,
  {Ishida}, T., {Ono}, T., {Tamura}, S., {Malasan}, H., {Yamasaki}, A.,
  {Hashimoto}, O., {Kambe}, E., {Takeuti}, M., {Kato}, T., {Kato}, M., {Chen},
  J.-S., {Leibowitz}, E.~M., {Roth}, M.~M., {Soffner}, T., \& {Mitsch}, W.
  1996, \aj, 112, 2699

\bibitem[{{Busso} {et~al.}(1999){Busso}, {Gallino}, \& {Wasserburg}}]{busso:99}
{Busso}, M., {Gallino}, R., \& {Wasserburg}, G.~J. 1999, ARA\&A, 37, 239

\bibitem[{{Cameron} \& {Fowler}(1971)}]{cameron:71}
{Cameron}, A.~G.~W. \& {Fowler}, W.~A. 1971, ApJ, 164, 111

\bibitem[{{Clayton} \& {De Marco}(1997)}]{clayton:97}
{Clayton}, G.~C. \& {De Marco}, O. 1997, AJ, 114, 2679

\bibitem[{C\'orsico \& Althaus(2005)}]{corsico:05}
C\'orsico, A.~H. \& Althaus, L.~G. 2005, A\&A, 439, L31

\bibitem[{{Cox}(2003)}]{cox:03}
{Cox}, A.~N. 2003, ApJ, 585, 975

\bibitem[{{Crowther} {et~al.}(1998){Crowther}, {de Marco}, \&
  {Barlow}}]{crowther:98}
{Crowther}, P.~A., {de Marco}, O., \& {Barlow}, M.~J. 1998, \mnras, 296, 367

\bibitem[{{De Marco} \& {Barlow}(2001)}]{demarco:01a}
{De Marco}, O. \& {Barlow}, M.~J. 2001, \apss, 275, 53

\bibitem[{{De Marco} {et~al.}(2003{\natexlab{a}}){De Marco}, Sandquist, {Mac
  Low}, Herwig, \& Taam}]{demarco:03a}
{De Marco}, O., Sandquist, E.~L., {Mac Low}, M., Herwig, F., \& Taam, R.~E.
  2003{\natexlab{a}}, in The eighth Texas-Mexico Conference on Astrophyiscs:
  Energetics of Cosmic Plasmas, 24

\bibitem[{{De Marco} {et~al.}(2003{\natexlab{b}}){De Marco}, {Sandquist}, {Mac
  Low}, {Herwig}, \& {Taam}}]{demarco:02b}
{De Marco}, O., {Sandquist}, E.~L., {Mac Low}, M.-M., {Herwig}, F., \& {Taam},
  R.~E. 2003{\natexlab{b}}, in Revista Mexicana de Astronomia y Astrofisica
  Conference Series, Vol.~15, 34

\bibitem[{{Denissenkov} \& {Herwig}(2004)}]{denissenkov:04a}
{Denissenkov}, P.~A. \& {Herwig}, F. 2004, \apj, 612, 1081

\bibitem[{{Denissenkov} \& {Tout}(2003)}]{denissenkov:02}
{Denissenkov}, P.~A. \& {Tout}, C.~A. 2003, MNRAS, 340, 722

\bibitem[{{Dimotakis}(2005)}]{dimotakis:05}
{Dimotakis}, P.~E. 2005, {Annu. Rev. Fluid Mech.}, 37, 329

\bibitem[{{Dreizler}(1998)}]{dreizler:98b}
{Dreizler}, S. 1998, Baltic Astronomy, 7, 71

\bibitem[{Dreizler \& Heber(1998)}]{dreizler:98a}
Dreizler, S. \& Heber, U. 1998, A\&A, 334, 618

\bibitem[{Dreizler {et~al.}(1995)Dreizler, Werner, \& Heber}]{dreizler:95}
Dreizler, S., Werner, K., \& Heber, U. 1995, in White Dwarfs, ed. D.~Koester \&
  K.~Werner, LNP No. 443 (Heidelberg: Springer), 160

\bibitem[{{Duerbeck} \& {Benetti}(1996)}]{duerbeck:96}
{Duerbeck}, H.~W. \& {Benetti}, S. 1996, ApJ Lett., 468, L111

\bibitem[{Duerbeck {et~al.}(1997)Duerbeck, Benetti, Gautchy, van Genderen,
  Kemper, Lillier, \& Thomas}]{duerbeck:97}
Duerbeck, H.~W., Benetti, S., Gautchy, A., van Genderen, A.~M., Kemper, C.,
  Lillier, W., \& Thomas, T. 1997, AJ, 114, 1657

\bibitem[{{Duerbeck} {et~al.}(2000){Duerbeck}, {Liller}, {Sterken}, {Benetti},
  {van Genderen}, {Arts}, {Kurk}, {Janson}, {Voskes}, {Brogt}, {Arentoft}, {van
  der Meer}, \& {Dijkstra}}]{duerbeck:00}
{Duerbeck}, H.~W., {Liller}, W., {Sterken}, C., {Benetti}, S., {van Genderen},
  A.~M., {Arts}, J., {Kurk}, J.~D., {Janson}, M., {Voskes}, T., {Brogt}, E.,
  {Arentoft}, T., {van der Meer}, A., \& {Dijkstra}, R. 2000, AJ, 119, 2360

\bibitem[{Fedrow {et~al.}(2005)Fedrow, Clayton, Crowther, \&
  Kerber}]{fedrow:05}
Fedrow, J., Clayton, G., Crowther, P., \& Kerber, F. 2005, AAS meeting 207,
  \#182.20

\bibitem[{Fogel {et~al.}(2003)Fogel, Marco, \& Jacoby}]{fogel:03}
Fogel, J., Marco, O.~D., \& Jacoby, G. 2003, in Planetary Nebulae: Their
  Evolution and Role in the Universe, ed. S.~Kwok, M.~Dopita, \& R.~Sutherland,
  ASP, IAU Symp.\ 209, 235

\bibitem[{Freytag {et~al.}(1996)Freytag, Ludwig, \& Steffen}]{freytag:96}
Freytag, B., Ludwig, H.-G., \& Steffen, M. 1996, A\&A, 313, 497

\bibitem[{{Fu} \& Vauclair(2006)}]{fu:06}
{Fu}, J.-N. \& Vauclair, G. 2006, A\&A, submitted.

\bibitem[{Fujimoto(1977)}]{fujimoto:77}
Fujimoto, M.~Y. 1977, PASJ, 29, 331

\bibitem[{Gallino {et~al.}(1998)Gallino, Arlandini, Busso, Lugaro, Travaglio,
  Straniero, Chieffi, \& Limongi}]{gallino:97b}
Gallino, R., Arlandini, C., Busso, M., Lugaro, M., Travaglio, C., Straniero,
  O., Chieffi, A., \& Limongi, M. 1998, ApJ, 497, 388

\bibitem[{{Garcia Lopez} \& {Spruit}(1991)}]{garcia-lopez:91}
{Garcia Lopez}, R.~J. \& {Spruit}, H.~C. 1991, ApJ, 377, 268

\bibitem[{{Gautschy}(1997)}]{gautschy:97}
{Gautschy}, A. 1997, \aap, 320, 811

\bibitem[{{Gautschy} {et~al.}(2005){Gautschy}, {Althaus}, \&
  {Saio}}]{gautschy:05}
{Gautschy}, A., {Althaus}, L.~G., \& {Saio}, H. 2005, \aap, 438, 1013

\bibitem[{{Gehrz} {et~al.}(2005){Gehrz}, {Woodward}, {Temim}, {Lyke}, \&
  {Mason}}]{gehrz:05}
{Gehrz}, R.~D., {Woodward}, C.~E., {Temim}, T., {Lyke}, J.~E., \& {Mason},
  C.~G. 2005, ApJ, 623, 1105

\bibitem[{{Gonzalez} {et~al.}(1998){Gonzalez}, {Lambert}, {Wallerstein}, {Rao},
  {Smith}, \& {McCarthy}}]{gonzalez:98}
{Gonzalez}, G., {Lambert}, D.~L., {Wallerstein}, G., {Rao}, N.~K., {Smith},
  V.~V., \& {McCarthy}, J.~K. 1998, ApJS, 114, 133

\bibitem[{{G{\'o}rny}(2001)}]{gorny:01}
{G{\'o}rny}, S.~K. 2001, \apss, 275, 67

\bibitem[{{Gr{\"a}fener} {et~al.}(2003){Gr{\"a}fener}, {Hamann}, \&
  {Pe{\~n}a}}]{graefener:03}
{Gr{\"a}fener}, G., {Hamann}, W.-R., \& {Pe{\~n}a}, M. 2003, in Planetary
  Nebulae: Their Evolution and Role in the Universe, ed. S.~Kwok, M.~Dopita, \&
  R.~Sutherland, ASP, IAU Symp.\ 209, 577

\bibitem[{Grevesse \& Sauval(2001)}]{grevesse:01}
Grevesse, N. \& Sauval, A.~J. 2001, in Encyclopedia of Astronomy and
  Astrophysics (IOP Publisheing Ltd. and Nature Publishing Group), 2453

\bibitem[{{Hajduk} {et~al.}(2005){Hajduk}, {Zijlstra}, {Herwig}, {van Hoof},
  {Kerber}, {Kimeswenger}, {Pollacco}, {Evans}, {Lop{\'e}z}, {Bryce}, {Eyres},
  \& {Matsuura}}]{hajduk:05}
{Hajduk}, M., {Zijlstra}, A.~A., {Herwig}, F., {van Hoof}, P.~A.~M., {Kerber},
  F., {Kimeswenger}, S., {Pollacco}, D.~L., {Evans}, A., {Lop{\'e}z}, J.~A.,
  {Bryce}, M., {Eyres}, S.~P.~S., \& {Matsuura}, M. 2005, Science, 308, 231

\bibitem[{Hamann(1997)}]{hamann:97}
Hamann, W.-R. 1997, in Planetary Nebulae, IAU Symp.\,180, ed. H.~J. Habing \&
  H.~J. G. L.~M. Lamers (Kluwer), 91

\bibitem[{Hamann(2003)}]{hamann:03a}
Hamann, W.-R. 2003, in Planetary Nebulae: Their Evolution and Role in the
  Universe, ed. S.~Kwok, M.~Dopita, \& R.~Sutherland, ASP, IAU Symp.\ 209, 203

\bibitem[{{Hamann} {et~al.}(2003){Hamann}, {Pe{\~n}a}, {Gr{\"a}fener}, \&
  {Ruiz}}]{hamann:03b}
{Hamann}, W.-R., {Pe{\~n}a}, M., {Gr{\"a}fener}, G., \& {Ruiz}, M.~T. 2003,
  \aap, 409, 969

\bibitem[{Hamann {et~al.}(2005)Hamann, Todt, \& Gr\"afener}]{hamann:05a}
Hamann, W.-R., Todt, H., \& Gr\"afener, G. 2005, in AIP Conference Proceedings,
  Vol. 804, Planetary Nebulae as Astronomical Tools, ed. R.~Szczerba,
  G.~Stasi\'nska, \& S.~K. G\'orny, 153

\bibitem[{{Heap}(1975)}]{heap:75}
{Heap}, S.~R. 1975, \apj, 196, 195

\bibitem[{Herald {et~al.}(2005)Herald, Bianchi, \& Hillier}]{herald:05}
Herald, J.~E., Bianchi, L., \& Hillier, D.~J. 2005, ApJ, 627, 424

\bibitem[{Herwig(2000)}]{herwig:99a}
Herwig, F. 2000, A\&A, 360, 952

\bibitem[{Herwig(2001)}]{herwig:01a}
---. 2001, ApJ Lett., 554, L71

\bibitem[{{Herwig}(2001)}]{herwig:00a}
{Herwig}, F. 2001, \apss, 275, 15

\bibitem[{{Herwig}(2003)}]{2003IAUS..209..111H}
{Herwig}, F. 2003, in Planetary Nebulae: Their Evolution and Role in the
  Universe, ed. S.~Kwok, M.~Dopita, \& R.~Sutherland, ASP, IAU Symp.\ 209, 111

\bibitem[{Herwig(2005)}]{herwig:04c}
Herwig, F. 2005, ARA\&A, 43, 435

\bibitem[{Herwig {et~al.}(1999)Herwig, Bl\"ocker, Langer, \&
  Driebe}]{herwig:99c}
Herwig, F., Bl\"ocker, T., Langer, N., \& Driebe, T. 1999, A\&A, 349, L5

\bibitem[{Herwig {et~al.}(1997)Herwig, Bl\"ocker, Sch\"onberner, \& {El
  Eid}}]{herwig:97}
Herwig, F., Bl\"ocker, T., Sch\"onberner, D., \& {El Eid}, M.~F. 1997, A\&A,
  324, L81

\bibitem[{Herwig {et~al.}(2006)Herwig, Freytag, Hueckstaedt, \&
  Timmes}]{herwig:06a}
Herwig, F., Freytag, B., Hueckstaedt, R.~M., \& Timmes, F.~X. 2006, ApJ, in
  press

\bibitem[{{Herwig} \& {Langer}(2001)}]{herwig:00f}
{Herwig}, F. \& {Langer}, N. 2001, Nuclear Physics A, 688, 221

\bibitem[{{Herwig} {et~al.}(2003{\natexlab{a}}){Herwig}, {Langer}, \&
  {Lugaro}}]{herwig:02a}
{Herwig}, F., {Langer}, N., \& {Lugaro}, M. 2003{\natexlab{a}}, ApJ, 593, 1056

\bibitem[{{Herwig} {et~al.}(2003{\natexlab{b}}){Herwig}, {Lugaro}, \&
  {Werner}}]{2003IAUS..209...85H}
{Herwig}, F., {Lugaro}, M., \& {Werner}, K. 2003{\natexlab{b}}, in Planetary
  Nebulae: Their Evolution and Role in the Universe, ed. S.~Kwok, M.~Dopita, \&
  R.~Sutherland, ASP, IAU Symp.\ 209, 85

\bibitem[{Herwig {et~al.}(1998)Herwig, Sch\"onberner, \&
  Bl\"ocker}]{herwig:98b}
Herwig, F., Sch\"onberner, D., \& Bl\"ocker, T. 1998, A\&A, 340, L43

\bibitem[{H{\"u}gelmeyer {et~al.}(2006)H{\"u}gelmeyer, Dreizler, Homeier,
  Krzesi{\'n}ski, Werner, \& {et al.}}]{huegelmeyer:06}
H{\"u}gelmeyer, S.~D., Dreizler, S., Homeier, D., Krzesi{\'n}ski, J., Werner,
  K., \& {et al.} 2006, A\&A, submitted.

\bibitem[{{H{\"u}gelmeyer} {et~al.}(2005){H{\"u}gelmeyer}, {Dreizler},
  {Werner}, {Krzesi{\'n}ski}, {Nitta}, \& {Kleinman}}]{huegelmeyer:05}
{H{\"u}gelmeyer}, S.~D., {Dreizler}, S., {Werner}, K., {Krzesi{\'n}ski}, J.,
  {Nitta}, A., \& {Kleinman}, S.~J. 2005, A\&A, 442, 309

\bibitem[{Iben(1995)}]{iben:95}
Iben, Jr., I. 1995, Physics Reports, 250, 1

\bibitem[{Iben {et~al.}(1983)Iben, Kaler, Truran, \& Renzini}]{iben:83a}
Iben, Jr., I., Kaler, J.~B., Truran, J.~W., \& Renzini, A. 1983, ApJ, 264, 605

\bibitem[{Iben \& MacDonald(1995)}]{iben:95b}
Iben, Jr., I. \& MacDonald, J. 1995, in LNP Vol. 443: White Dwarfs, ed.
  D.~Koester \& K.~Werner (Heidelberg: Springer), 48

\bibitem[{Iben \& Renzini(1983)}]{iben:83b}
Iben, Jr., I. \& Renzini, A. 1983, ARA\&A, 21, 271

\bibitem[{{Iwamoto} {et~al.}(2004){Iwamoto}, {Kajino}, {Mathews}, {Fujimoto},
  \& {Aoki}}]{iwamoto:04}
{Iwamoto}, N., {Kajino}, T., {Mathews}, G.~J., {Fujimoto}, M.~Y., \& {Aoki}, W.
  2004, ApJ, 602, 378

\bibitem[{Jahn(2005)}]{jahn:05}
Jahn, D. 2005, {Diploma Thesis}, University of T\"ubingen, Germany

\bibitem[{Jorissen {et~al.}(1992)Jorissen, Smith, \& Lambert}]{jorissen:92}
Jorissen, A., Smith, V.~V., \& Lambert, D.~L. 1992, A\&A, 261, 164

\bibitem[{{Kawaler} \& {Bradley}(1994)}]{kawaler:94}
{Kawaler}, S.~D. \& {Bradley}, P.~A. 1994, ApJ, 427, 415

\bibitem[{Kawaler {et~al.}(1995)Kawaler, O'Brien, Clemens, \& {et
  al.}}]{kawaler:95}
Kawaler, S.~D., O'Brien, M.~S., Clemens, J.~C., \& {et al.} 1995, ApJ, 350, 363

\bibitem[{Kawaler {et~al.}(2004)Kawaler, Potter, Vu\v{c}kovi\'c, \&
  et~al.}]{kawaler:04}
Kawaler, S.~D., Potter, E.~M., Vu\v{c}kovi\'c, M., \& et~al. 2004, A\&A, 428,
  969

\bibitem[{{Koesterke}(2001)}]{koesterke:01}
{Koesterke}, L. 2001, Ap\&SS, 275, 41

\bibitem[{Koesterke {et~al.}(1998)Koesterke, Dreizler, \& Rauch}]{koesterke:98}
Koesterke, L., Dreizler, S., \& Rauch, T. 1998, A\&A, 330, 1041

\bibitem[{Koesterke \& Hamann(1997)}]{koesterke:97}
Koesterke, L. \& Hamann, W.-R. 1997, in Planetary Nebulae, ed. H.~J. Habing \&
  H.~J. G. L.~M. Lamers, IAU Symp.\ 180 (Dordrecht: Kluwer), 114

\bibitem[{{Koesterke} \& {Hamann}(1997)}]{koesterke:97b}
{Koesterke}, L. \& {Hamann}, W.~R. 1997, A\&A, 320, 91

\bibitem[{Koesterke \& Werner(1998)}]{koesterke:98b}
Koesterke, L. \& Werner, K. 1998, ApJ Lett., 500, 55

\bibitem[{{Lawlor} \& {MacDonald}(2002)}]{lawlor:02}
{Lawlor}, T.~M. \& {MacDonald}, J. 2002, Ap\&SS, 279, 123

\bibitem[{{Lawlor} \& {MacDonald}(2003)}]{lawlor:03}
---. 2003, ApJ, 583, 913

\bibitem[{{Lechner} \& {Kimeswenger}(2004)}]{lechner:04}
{Lechner}, M.~F.~M. \& {Kimeswenger}, S. 2004, A\&A, 426, 145

\bibitem[{Leuenhagen \& Hamann(1998)}]{leuenhagen:98}
Leuenhagen, U. \& Hamann, W.-R. 1998, A\&A, 330, 265

\bibitem[{{Lugaro} {et~al.}(2003){Lugaro}, {Herwig}, {Lattanzio}, {Gallino}, \&
  {Straniero}}]{lugaro:02a}
{Lugaro}, M., {Herwig}, F., {Lattanzio}, J.~C., {Gallino}, R., \& {Straniero},
  O. 2003, ApJ, 586, 1305

\bibitem[{{Lugaro} {et~al.}(2004){Lugaro}, {Ugalde}, {Karakas}, {G{\" o}rres},
  {Wiescher}, {Lattanzio}, \& {Cannon}}]{lugaro:04a}
{Lugaro}, M., {Ugalde}, C., {Karakas}, A.~I., {G{\" o}rres}, J., {Wiescher},
  M., {Lattanzio}, J.~C., \& {Cannon}, R.~C. 2004, ApJ, 615, 934

\bibitem[{{Lugaro} {et~al.}(1999){Lugaro}, {Zinner}, {Gallino}, \&
  {Amari}}]{lugaro:99}
{Lugaro}, M., {Zinner}, E., {Gallino}, R., \& {Amari}, S. 1999, ApJ, 527, 369

\bibitem[{M\'endez(1991)}]{mendez:91}
M\'endez, R.~H. 1991, in Evolution of Stars: The Photospheric Abundance
  Connection, IAU Symp. 145, ed. G.~Michaud \& A.~Tutukov, 375

\bibitem[{{Metcalfe} {et~al.}(2005){Metcalfe}, {Nather}, {Watson}, {Kim},
  {Park}, \& {Handler}}]{metcalfe:05}
{Metcalfe}, T.~S., {Nather}, R.~E., {Watson}, T.~K., {Kim}, S.-L., {Park},
  B.-G., \& {Handler}, G. 2005, \aap, 435, 649

\bibitem[{{Miksa} {et~al.}(2002){Miksa}, {Deetjen}, {Dreizler}, {Kruk},
  {Rauch}, \& {Werner}}]{miksa:02}
{Miksa}, S., {Deetjen}, J.~L., {Dreizler}, S., {Kruk}, J.~W., {Rauch}, T., \&
  {Werner}, K. 2002, A\&A, 389, 953

\bibitem[{{Miller~Bertolami} \& Althaus(2006)}]{bertolami:06}
{Miller~Bertolami}, M.~M. \& Althaus, L.~G. 2006, A\&A, submitted

\bibitem[{{Miller~Bertolami} {et~al.}(2005){Miller~Bertolami}, Serenelli, \&
  Panei}]{miller-bertolami:05}
{Miller~Bertolami}, M.~M., Serenelli, L. G. A. A.~M., \& Panei, J.~A. 2005,
  A\&A, in press, astro-ph/0511406

\bibitem[{{Montalban}(1994)}]{montalban:94}
{Montalban}, J. 1994, A\&A, 281, 421

\bibitem[{Morgan {et~al.}(2003)Morgan, Parker, \& Cohen}]{morgan:03a}
Morgan, D.~H., Parker, Q.~A., \& Cohen, M. 2003, MNRAS, 346, 719

\bibitem[{Mowlavi(1999)}]{mowlavi:99}
Mowlavi, N. 1999, A\&A, 344, 617

\bibitem[{Mowlavi {et~al.}(1996)Mowlavi, Jorissen, \& Arnould}]{mowlavi:96}
Mowlavi, N., Jorissen, A., \& Arnould, M. 1996, A\&A, 311, 803

\bibitem[{Nagel {et~al.}(2006)Nagel, Schuh, Kusterer, Stahn, H{\"u}gelmeyer,
  Dreizler, G{\"a}nsicke, \& Schreiber}]{nagel:06}
Nagel, T., Schuh, S., Kusterer, D.-J., Stahn, T., H{\"u}gelmeyer, S.~D.,
  Dreizler, S., G{\"a}nsicke, B., \& Schreiber, M. 2006, A\&A, submitted.

\bibitem[{{Napiwotzki}(1999)}]{napiwotzki:99b}
{Napiwotzki}, R. 1999, A\&A, 350, 101

\bibitem[{{Napiwotzki} \& {Sch\"onberner}(1995)}]{napiwotzki:95}
{Napiwotzki}, R. \& {Sch\"onberner}, D. 1995, A\&A, 301, 545

\bibitem[{Napiwotzki {et~al.}(1991)Napiwotzki, Sch\"onberner, \&
  Weidemann}]{napiwotzki:91}
Napiwotzki, R., Sch\"onberner, D., \& Weidemann, V. 1991, A\&A, 243, L5

\bibitem[{{Pandey} {et~al.}(2005){Pandey}, {Lambert}, {Jeffery}, \& {Kameswara
  Rao}}]{pandey:05}
{Pandey}, G., {Lambert}, D.~L., {Jeffery}, C.~S., \& {Kameswara Rao}, N. 2005,
  ApJ, in press, astro-ph/0510161

\bibitem[{{Press}(1981)}]{press:81}
{Press}, W.~H. 1981, ApJ, 245, 286

\bibitem[{Quirion {et~al.}(2004)Quirion, Fontaine, \& Brassard}]{quirion:04}
Quirion, P.~O., Fontaine, G., \& Brassard, P. 2004, ApJ, 610, 436

\bibitem[{Quirion {et~al.}(2005{\natexlab{a}})Quirion, Fontaine, \&
  Brassard}]{quirion:05a}
---. 2005{\natexlab{a}}, A\&A, 441, 231

\bibitem[{Quirion {et~al.}(2005{\natexlab{b}})Quirion, Fontaine, \&
  Brassard}]{quirion:05b}
---. 2005{\natexlab{b}}, Mem. Soc. Astron. Ital., 75, 282

\bibitem[{{Rauch} {et~al.}(1998){Rauch}, {Dreizler}, \& {Wolff}}]{rauch:98}
{Rauch}, T., {Dreizler}, S., \& {Wolff}, B. 1998, A\&A, 338, 651

\bibitem[{{Rauch} {et~al.}(1991){Rauch}, {Heber}, {Hunger}, {Werner}, \&
  {Neckel}}]{rauch:91}
{Rauch}, T., {Heber}, U., {Hunger}, K., {Werner}, K., \& {Neckel}, T. 1991,
  \aap, 241, 457

\bibitem[{{Rauch} \& {Werner}(1995)}]{rauch:95}
{Rauch}, T. \& {Werner}, K. 1995, in LNP Vol. 443: White Dwarfs, ed. D.~Koester
  \& K.~Werner (Heidelberg: Springer), 186

\bibitem[{Reiff {et~al.}(2006)Reiff, Jahn, Rauch, Werner, Kruk, \&
  Herwig}]{reiff:06a}
Reiff, E., Jahn, D., Rauch, T., Werner, K., Kruk, J.~W., \& Herwig, F. 2006, in
  ASP Conference Series, ed. C.~A. C.~Sterken, in press

\bibitem[{Reiff {et~al.}(2005)Reiff, Rauch, Werner, \& Kruk}]{reiff:05}
Reiff, E., Rauch, T., Werner, K., \& Kruk, J. 2005, in ASP Conf. Ser. 334:
  White Dwarfs, ed. D.~Koester \& S.~Moehler, 173

\bibitem[{Sackmann \& Boothroyd(1992)}]{sackmann:92}
Sackmann, I.-J. \& Boothroyd, A.~I. 1992, ApJ, 392, L71

\bibitem[{Saio(1996)}]{saio:96}
Saio, H. 1996, in ASP Conference Series, Vol.~96, Hydrogen-Deficient Stars, ed.
  U.~Heber \& C.~Jeffery, 361

\bibitem[{{Saio} \& {Jeffery}(2002)}]{saio:02}
{Saio}, H. \& {Jeffery}, C.~S. 2002, MNRAS, 333, 121

\bibitem[{{Scalo} {et~al.}(1975){Scalo}, {Despain}, \& {Ulrich}}]{scalo:75}
{Scalo}, J.~M., {Despain}, K.~H., \& {Ulrich}, R.~K. 1975, ApJ, 196, 805

\bibitem[{Sch\"onberner(1979)}]{schoenberner:79}
Sch\"onberner, D. 1979, A\&A, 79, 108

\bibitem[{Sch\"onberner(1983)}]{schoenberner:83}
---. 1983, ApJ, 272, 708

\bibitem[{{Sch{\"o}nberner} \& {Jeffery}(2002)}]{schoenberner:02}
{Sch{\"o}nberner}, D. \& {Jeffery}, C.~S. 2002, in ASP Conf. Ser. 279: Exotic
  Stars as Challenges to Evolution, IAU Coll. 279, ed. C.~A. Tout \& W.~{Van
  Hamme}, 173

\bibitem[{{Schwarzschild} \& {H{\"a}rm}(1967)}]{schwarzschild:67}
{Schwarzschild}, M. \& {H{\"a}rm}, R. 1967, \apj, 150, 961

\bibitem[{{Seitter}(1987{\natexlab{a}})}]{seitter:87a}
{Seitter}, W.~C. 1987{\natexlab{a}}, Mitteilungen der Astronomischen
  Gesellschaft Hamburg, 68, 244

\bibitem[{{Seitter}(1987{\natexlab{b}})}]{seitter:87b}
---. 1987{\natexlab{b}}, The Messenger, 50, 14

\bibitem[{{Siess} {et~al.}(2004){Siess}, {Goriely}, \& {Langer}}]{siess:04}
{Siess}, L., {Goriely}, S., \& {Langer}, N. 2004, A\&A, 415, 1089

\bibitem[{Smith \& Aller(1969)}]{smith:69}
Smith, L.~F. \& Aller, L.~H. 1969, ApJ, 157, 1245

\bibitem[{Starrfield {et~al.}(1983)Starrfield, Cox, Hodson, \&
  Pesnell}]{starrfield:83}
Starrfield, S., Cox, A.~N., Hodson, S.~W., \& Pesnell, W.~D. 1983, ApJ Lett.,
  268, 27

\bibitem[{Straniero {et~al.}(1997)Straniero, Chieffi, Limongi, Busso, Gallino,
  \& Arlandini}]{straniero:97}
Straniero, O., Chieffi, A., Limongi, M., Busso, M., Gallino, R., \& Arlandini,
  C. 1997, ApJ, 478, 332

\bibitem[{Tylenda {et~al.}(1993)Tylenda, Acker, \& Stenholm}]{tylenda:93}
Tylenda, R., Acker, A., \& Stenholm, B. 1993, A\&AS, 102, 595

\bibitem[{Vauclair {et~al.}(2002)Vauclair, Moskalik, Pfeiffer, \&
  et~al.}]{vauclair:02a}
Vauclair, G., Moskalik, P., Pfeiffer, B., \& et~al. 2002, A\&A, 381, 122

\bibitem[{{Werner}(1992)}]{werner:92b}
{Werner}, K. 1992, in LNP Vol.~401: The Atmospheres of Early-Type Stars, 401,
  273

\bibitem[{{Werner}(1996)}]{werner:96}
---. 1996, \aap, 309, 861

\bibitem[{{Werner} {et~al.}(2003){Werner}, {Deetjen}, {Dreizler}, {Rauch}, \&
  {Kruk}}]{werner:03}
{Werner}, K., {Deetjen}, J.~L., {Dreizler}, S., {Rauch}, T., \& {Kruk}, J.~W.
  2003, in IAU Symposium, ed. S.~Kwok, M.~Dopita, \& R.~Sutherland, ASP, IAU
  Symp.\ 209, 169

\bibitem[{{Werner} {et~al.}(1995){Werner}, {Dreizler}, {Heber}, {Rauch},
  {Wisotzki}, \& {Hagen}}]{werner:95}
{Werner}, K., {Dreizler}, S., {Heber}, U., {Rauch}, T., {Wisotzki}, L., \&
  {Hagen}, H.-J. 1995, \aap, 293, L75

\bibitem[{Werner {et~al.}(1992)Werner, Hamann, Heber, Napiwotzki, Rauch, \&
  Wessolowski}]{werner:92c}
Werner, K., Hamann, W.-R., Heber, U., Napiwotzki, R., Rauch, T., \&
  Wessolowski, U. 1992, A\&A, 259, L69

\bibitem[{{Werner} \& {Heber}(1991)}]{werner:91b}
{Werner}, K. \& {Heber}, U. 1991, \aap, 247, 476

\bibitem[{Werner {et~al.}(1994)Werner, Heber, \& Fleming}]{werner:94b}
Werner, K., Heber, U., \& Fleming, T. 1994, A\&A, 284, 907

\bibitem[{{Werner} {et~al.}(1991){Werner}, {Heber}, \& {Hunger}}]{werner:91}
{Werner}, K., {Heber}, U., \& {Hunger}, K. 1991, A\&A, 244, 437

\bibitem[{{Werner} \& {Koesterke}(1992)}]{werner:92}
{Werner}, K. \& {Koesterke}, L. 1992, in LNP Vol.~401: The Atmospheres of
  Early-Type Stars, 401, 288

\bibitem[{{Werner} \& {Rauch}(1994)}]{werner:94}
{Werner}, K. \& {Rauch}, T. 1994, A\&A, 284, L5

\bibitem[{{Werner} {et~al.}(2004{\natexlab{a}}){Werner}, {Rauch}, {Barstow}, \&
  {Kruk}}]{werner:04b}
{Werner}, K., {Rauch}, T., {Barstow}, M.~A., \& {Kruk}, J.~W.
  2004{\natexlab{a}}, A\&A, 421, 1169

\bibitem[{{Werner} {et~al.}(2005){Werner}, {Rauch}, \& {Kruk}}]{werner:05b}
{Werner}, K., {Rauch}, T., \& {Kruk}, J.~W. 2005, A\&A, 433, 641

\bibitem[{{Werner} {et~al.}(2004{\natexlab{b}}){Werner}, {Rauch}, {Napiwotzki},
  {Christlieb}, {Reimers}, \& {Karl}}]{werner:04c}
{Werner}, K., {Rauch}, T., {Napiwotzki}, R., {Christlieb}, N., {Reimers}, D.,
  \& {Karl}, C.~A. 2004{\natexlab{b}}, A\&A, 424, 657

\bibitem[{{Werner} {et~al.}(2004{\natexlab{c}}){Werner}, {Rauch}, {Reiff},
  {Kruk}, \& {Napiwotzki}}]{werner:04}
{Werner}, K., {Rauch}, T., {Reiff}, E., {Kruk}, J.~W., \& {Napiwotzki}, R.
  2004{\natexlab{c}}, A\&A, 427, 685

\bibitem[{{Wesemael} {et~al.}(1985){Wesemael}, {Green}, \&
  {Liebert}}]{wesemael:85}
{Wesemael}, F., {Green}, R.~F., \& {Liebert}, J. 1985, ApJS, 58, 379

\bibitem[{Wood \& Faulkner(1986)}]{wood:86}
Wood, P.~R. \& Faulkner, D.~J. 1986, ApJ, 307, 659

\bibitem[{{Young} {et~al.}(2005){Young}, {Meakin}, {Arnett}, \&
  {Fryer}}]{young:05b}
{Young}, P.~A., {Meakin}, C., {Arnett}, D., \& {Fryer}, C.~L. 2005, ApJ Lett.,
  629, L101

\bibitem[{Zahn(1991)}]{zahn:91}
Zahn, J.-P. 1991, A\&A, 252, 191

\end{thebibliography}

\end{document}